\documentclass[aps,amsmath,amsfonts,nofootinbib,preprintnumbers, superscriptaddress,eqsecnum,secnumarabic]{revtex4}
\usepackage{graphicx,slashed,hyperref}
\pdfoutput=1
\usepackage[utf8]{inputenc}
\hypersetup{
   bookmarks=true,         
   unicode=true,          
   pdftoolbar=true,        
   pdfmenubar=true,        
   pdffitwindow=false,     
   pdfstartview={FitH},    
   pdftitle={My title},    
   pdfauthor={Author},     
   pdfsubject={Subject},   
   pdfcreator={Creator},   
   pdfproducer={Producer}, 
   pdfkeywords={keyword1} {key2} {key3}, 
   pdfnewwindow=true,      
   colorlinks=true,       
   linkcolor=blue,          
   citecolor=magenta,        
   filecolor=magenta,      
   urlcolor=blue           
}
\usepackage{amsmath}
\usepackage{amssymb}
\usepackage{xcolor}

\usepackage{comment}
\newcommand{\be}{\begin{equation}}
\newcommand{\ee}{\end{equation}}
\def\bsp#1\esp{\begin{split}#1\end{split}}

\newcommand{\Eq}[1]{Eq.~(\ref{#1})}
\newcommand{\Fig}[1]{Fig.~\ref{#1}}
\newcommand{\Sec}[1]{Section~\ref{#1}}

\newcommand\ion[2]{\text{#1\,\textsc{\lowercase{#2}}}}  

\begin{document}

\title{Variations in fundamental constants at the cosmic dawn}
\author{Laura~Lopez-Honorez}
\email{llopezho@ulb.ac.be}
\affiliation{Service de Physique Th\'eorique, CP225, Universit\'e Libre de Bruxelles, Bld du Triomphe, 1050 Brussels; and Vrije Universiteit Brussel and The International Solvay Institutes, Pleinlaan 2, 1050 Brussels, Belgium.}
\author{Olga Mena}
\email{olga.mena@ific.uv.es}
\author{Sergio Palomares-Ruiz}
\email{sergiopr@ific.uv.es}
\author{Pablo Villanueva-Domingo}
\email{pablo.villanueva@ific.uv.es}
\author{Samuel J. Witte}
\email{sam.witte@ific.uv.es}
\affiliation{Instituto de F\'isica Corpuscular (IFIC),
  CSIC-Universitat de Valencia,\\
Apartado de Correos 22085,  E-46071, Spain}

\preprint{ULB-TH/20-03}

\begin{abstract}
The observation of space-time variations in fundamental constants would provide strong evidence for the existence of new light degrees of freedom in the theory of Nature. Robustly constraining such scenarios requires exploiting observations that span different scales and probe the state of the Universe at different epochs. In the context of cosmology, both the cosmic microwave background and the Lyman-$\alpha$ forest have proven to be powerful tools capable of constraining variations in electromagnetism, however at the moment there do not exist cosmological probes capable of bridging the gap between recombination and reionization. In the near future, radio telescopes will attempt to measure the 21cm transition of neutral hydrogen during the epochs of reionization and the cosmic dawn (and potentially the tail end of the dark ages); being inherently sensitive to electromagnetic phenomena, these experiments will offer a unique perspective on space-time variations of the fine-structure constant and the electron mass. We show here that large variations in these fundamental constants would produce features on the 21cm power spectrum that may be distinguishable from astrophysical uncertainties. Furthermore, we forecast the sensitivity for the Square Kilometer Array, and show that the 21cm power spectrum may be able to constrain variations at the level of ${\cal O}(10^{-3})$.
\end{abstract}

\maketitle

\section{Introduction}

Space-time variations of fundamental constants of Nature, such as the fine-structure constant or the electron mass, offer a strong test for extensions to the standard cosmological and elementary particle models, such as extra dimensions or modified gravity~\cite{Uzan:2002vq, Uzan:2010pm, Martins:2017yxk}. Consequently, searches of space-time variations in the observed values of these fundamental constants can provide hints of new and well-motivated exotic physics scenarios.  It is therefore interesting to test possible deviations from their current values at both large and small distances, as well as across a wide array of cosmological times (redshifts).  

Focusing on the fine-structure constant $\alpha$, which provides the strength of the photon-charged lepton interactions, and on the electron mass $m_e$, limits have been derived at low redshifts $z \lesssim 6 $ using, for instance, absorption lines in the spectra of distant quasars (e.g., the Lyman-$\alpha$ forest)~\cite{Dzuba:1999zz, Ivanchik:2001ji, Murphy:2002kik, Webb:2010hc, King:2012id, Kotus:2016xxb, Martins:2017yxk, Murphy:2017xaz, Leite:2018cgh, Leite:2019txc, Hart:2019gvj, Wilczynska:2020rxx}, and at $z\simeq1000$ (at the period of recombination) from observations of the cosmic microwave background (CMB) ~\cite{Kaplinghat:1998ry, Avelino:2000ea, Battye:2000ds, Avelino:2001nr, Martins:2002iv, Martins:2003pe, Rocha:2003gc, Scoccola:2008jw, Nakashima:2008cb, Scoccola:2009iz, Landau:2010zs, Menegoni:2012tq, Ade:2014zfo, Hart:2017ndk}. The tightest constraints to date at the time of recombination on $\alpha$ and $m_e$ are at the level of $\alpha/\alpha_ 0 - 1 = (- 0.7 \pm 2) \times 10^{-3}$ and $m_e/m_{e0} - 1 = (3.9 \pm 7.4) \times 10^{-3}$~\cite{Hart:2017ndk}, where the subscript $0$ refers to their current local value. The low-redshift constraints using atomic and molecular absorption lines, however, are two-to-three orders of magnitude tighter~\cite{Webb:1998cq, Dzuba:1999zz}. The most stringent of these relies on the observation of methanol absorption lines in the PKS1830-211 lensing galaxy at $z=0.89$, and probes the proton-to-electron mass ratio  $\mu \equiv m_p/m_e$ at the level of  $\mu/\mu_0 - 1 = (-1.0 \pm 0.8_{\rm {stat}}\pm 1.0_{\rm {sys}}) \times 10^{-7}$~\cite{Bagdonaite:2013sia}. However, other observations have comparable sensitivity~\cite{Webb:2010hc, King:2012id, Kotus:2016xxb, Martins:2017yxk, Wilczynska:2020rxx}.

21cm cosmology offers a unique test at redshifts higher than the reionization period $z \gtrsim 6$, during the so-called dark ages and the period of the cosmic dawn, when the first stars started to form. These measurements will help constraining variations of fundamental constants in a completely uncharted era~\cite{Khatri:2007yv, Flambaum:2010um, Khatri:2010ns}. While it is unlikely that near-future constraints on the time variations of $\alpha$ and $m_e$ from measurements of the 21cm power spectrum, by for instance the Square Kilometer Array (SKA)~\cite{Mellema:2012ht, Bull:2018lat}, will be as stringent as those obtained using measurements of the quasar absorption spectra, they are valuable, since they will offer a \emph{unique} probe across a wide range of redshift (i.e., $6 < z < 30$). Furthermore, the large number of available redshifts could allow for analyses of the putative time dependence of both $\alpha$ and $m_e$, in contrast to CMB measurements which probe a narrow shell around $z \sim 1100$. Notice that previous works~\cite{Khatri:2007yv, Flambaum:2010um, Khatri:2010ns} studied the impact of the variation of $\alpha$ on the 21cm signal at $z > 30$, when collisional coupling plays an important role. Here, complementarily, we study the effect that changes in $\alpha$ and $m_e$ would have on the 21cm signal for $z < 20$, when the Lyman-$\alpha$ coupling plays a major role. Of significant importance in our analysis is the presence of degeneracies between astrophysical parameters and the variation of fundamental constants, which we carefully take into account.

It is therefore timely to compute the expected sensitivities from future giant radio telescope arrays targeting the 21cm signal to variations of fundamental constants in the \textit{era incognita}, that is, between the CMB and the reionization period. The structure of the paper is as follows. We start in \Sec{sec:21cm} by briefly describing the physics of 21cm cosmology in general. We then describe the implicit functional dependence of each relevant quantity on both, the fine-structure constant and the electron mass. In \Sec{sec:signal} we describe the effects of varying $\alpha$ and $m_e$ on the globally averaged differential brightness temperature and on the 21cm power spectrum. We present forecasts in \Sec{sec:forecasts} for SKA, allowing for changes in both $\alpha$ and $m_e$.  Finally, we summarize our results and conclude in \Sec{sec:concl}.

\section{21cm cosmology and its dependence on $\alpha$ and $m_e$}
\label{sec:21cm}

We begin by introducing the ingredients necessary to compute 21cm observables. These are obtained by solving the radiative transfer equation (in the absence of scattering) for radio waves traversing the IGM (readers interested in a more in-depth derivation could see, e.g., Refs.~\cite{Lewis:2007kz, Furlanetto:2006jb}). 

The brightness of a patch of neutral hydrogen (HI) relative to the CMB at a given redshift $z$ is expressed in terms of the differential brightness temperature,
\begin{equation}
\delta T_b(\nu) = \frac{T_{\rm{S}} - T_{\rm{CMB}}}{1 + z} \, \left(1 - e^{-\tau_{\nu_{21}}}\right) \simeq \frac{T_{\rm S} - T_{\rm{CMB}}}{1 + z} \, \tau_{\nu_{21}}  ~,
\label{eq:Tb}
\end{equation}
where $\tau_{\nu_{21}}$ is the optical depth of the 21cm line to the intergalactic medium (IGM), and $T_{\rm{S}}$ is the so-called `spin temperature', which is defined in such a way to measure the relative occupation of the ground and first excited states of neutral hydrogen. In the second line of \Eq{eq:Tb} we have explicitly assumed that the optical depth $\tau_{\nu_{21}}$ is small, a valid assumption across all redshifts of interest~\cite{Madau:1996cs, Furlanetto:2006jb, Pritchard:2011xb, Furlanetto:2015apc}.  The spin temperature $T_{\rm S}$ can be obtained by solving an equilibrium equation relating the excitation and de-excitation rates of neutral hydrogen; the solution, accounting for the possibility of spontaneous and stimulated excitation/de-excitation, collisional excitation/de-excitation, an indirect excitations via scattering with ambient Lyman-$\alpha$ photons, is given by~\cite{Hirata:2005mz},
\begin{equation}
T_{\rm S}^{-1} = \frac{T_{\textrm{CMB}}^{-1} + x_{\rm c} \, T_k^{-1} + x_\alpha \, T_{\rm c}^{-1} }{1 + x_{\rm c} + x_\alpha}  ~, 
\label{eq:TgamTsappr}
\end{equation}
where $x_{\rm c}$ and $x_\alpha$ are the coupling coefficients determining the importance of collisional and Lyman-$\alpha$ scattering processes, and $T_k$ and $T_c$ are the gas temperature and the so-called `color temperature', which are nearly equal for all environments of interest in this work. The resonant scattering of Lyman-$\alpha$ photons can induce an indirect transition between the hyperfine levels, a process dubbed the Wouthuysen-Field effect~\cite{Wouthuysen:1952, Field:1958}, and becomes important when the Lyman-$\alpha$ flux produced by the first sources permeates the medium, typically occurring around $z \sim 25$. This effect is of particular relevance for our study as it controls the value of the spin temperature for much of the cosmic dawn. In contrast, the collisional coupling $x_{\rm c}$, characterizing the efficiency of spin flips induced from collisions of neutral hydrogen, is only important at high redshifts $z \gtrsim 30$ or in extremely large over-densities, and could be safely neglected in \Eq{eq:TgamTsappr} for the redshifts of interest.\footnote{Although $x_c$ is included in the 21cm signal calculation by the public code used here (\texttt{21cmFASTv2}~\cite{Mesinger:2007pd, Mesinger:2010ne, Park:2018ljd}), we do not incorporate its dependence on $\alpha$ and $m_e$, as this would have no significant effect for the relevant range of redshifts (see Refs.~\cite{Khatri:2007yv, Flambaum:2010um, Khatri:2010ns} for  studies of the $\alpha$-dependence of the collisional de-excitation rate for $z > 30$). }

In what follows, we shall describe the explicit dependencies of the 21cm cosmological observables on the fundamental constants $\alpha$ and $m_e$. In order to maintain the highest level of clarity, from this point on we will explicitly write the functional dependence of any variable on these two quantities, and clearly state and justify when functions can be treated as approximately independent. For the sake of simplicity, we define the relative variations with respect to the local value of the fundamental constants as
\begin{equation}
\delta \alpha = \frac{\alpha(z)}{\alpha_0} - 1\quad {\rm and } \quad \delta m_e = \frac{m_{e}(z)}{m_{e0}} -1  ~.
\end{equation}
where the $0$ subscript refers to local (laboratory) value. We will assume throughout this work that variations are fixed over redshifts $6 \leq z \lesssim 1000$.\footnote{Note that, despite focusing on observations in the interval $6\lesssim z\lesssim30$, the initial conditions for the 21cm signal depend on the details of recombination. We have chosen here to include these effects self-consistently all the way through recombination making use of the public code {\tt Recfast++} to set these initial conditions (see Section~\ref{sec:forecasts}). Notice, though, that this is not technically necessary and does not have profound implications for our study.} Let us comment briefly, however, on the implications of dropping this assumption.

21cm experiments are expected to be able to accurately measure the  power spectrum for Fourier modes in the range $0.1 \, {\rm Mpc}^{-1} \lesssim k \lesssim 1 \, {\rm Mpc}^{-1}$, and in redshift bins of width $\Delta z \ll 1$. The limit adopted of a constant spatial and temporal shift in $\alpha$ and $m_e$ corresponds to the limit where temporal variations occur on scales larger than the maximum observable time, and spatial variations occur on scales not observable by the experiment (i.e., they can be larger than the maximum scale, in which case one is only sensitive to that value, or smaller than the minimum, in which case the effects average out). If temporal variations occur on scales smaller than the minimum observable timescale, the effects would again wash out and the experiment would have no sensitivity (however, we emphasize that one of the advantages that cosmological probes have is their sensitivity to large scales, so it is quite possible that complimentary astrophysical or laboratory experiments gain sensitivity to more rapidly varying signals). If space-time variations are such that 21cm experiments can resolve their dependence (i.e., they have temporal or spatial variations within the observable window), one would expect distinguishing features to appear that could strongly break any degeneracy with astrophysical modeling. While this would be an interesting observation, the sensitivity of an experiment to such variations is likely strongly dependent upon the adopted model. Our choice of space-time independent variations is thus general, practical, and to a large degree conservative.

\begin{figure*}
	\begin{centering}
		\includegraphics[width=.49\textwidth]{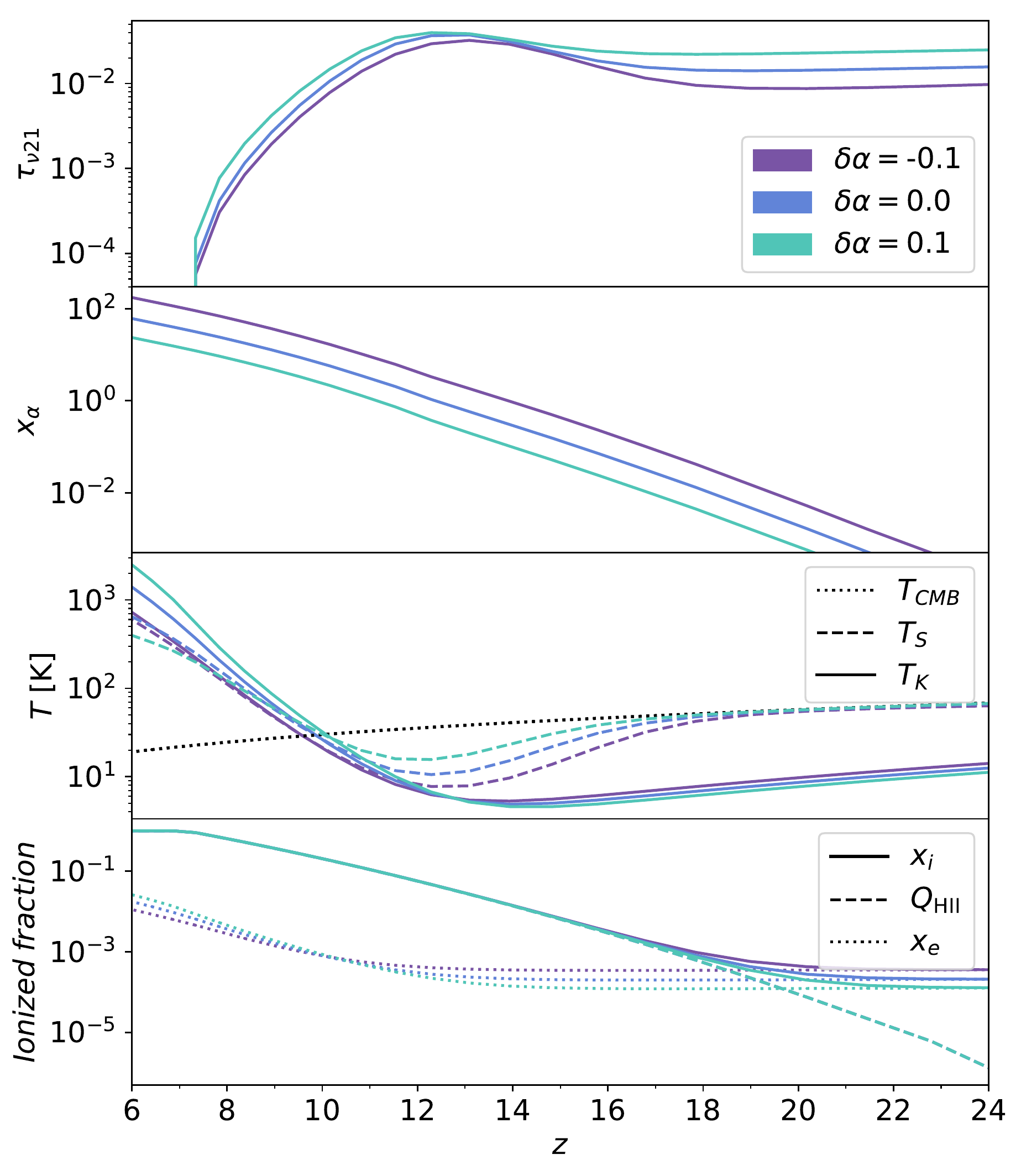}\hspace{1ex}
		\includegraphics[width=.49\textwidth]{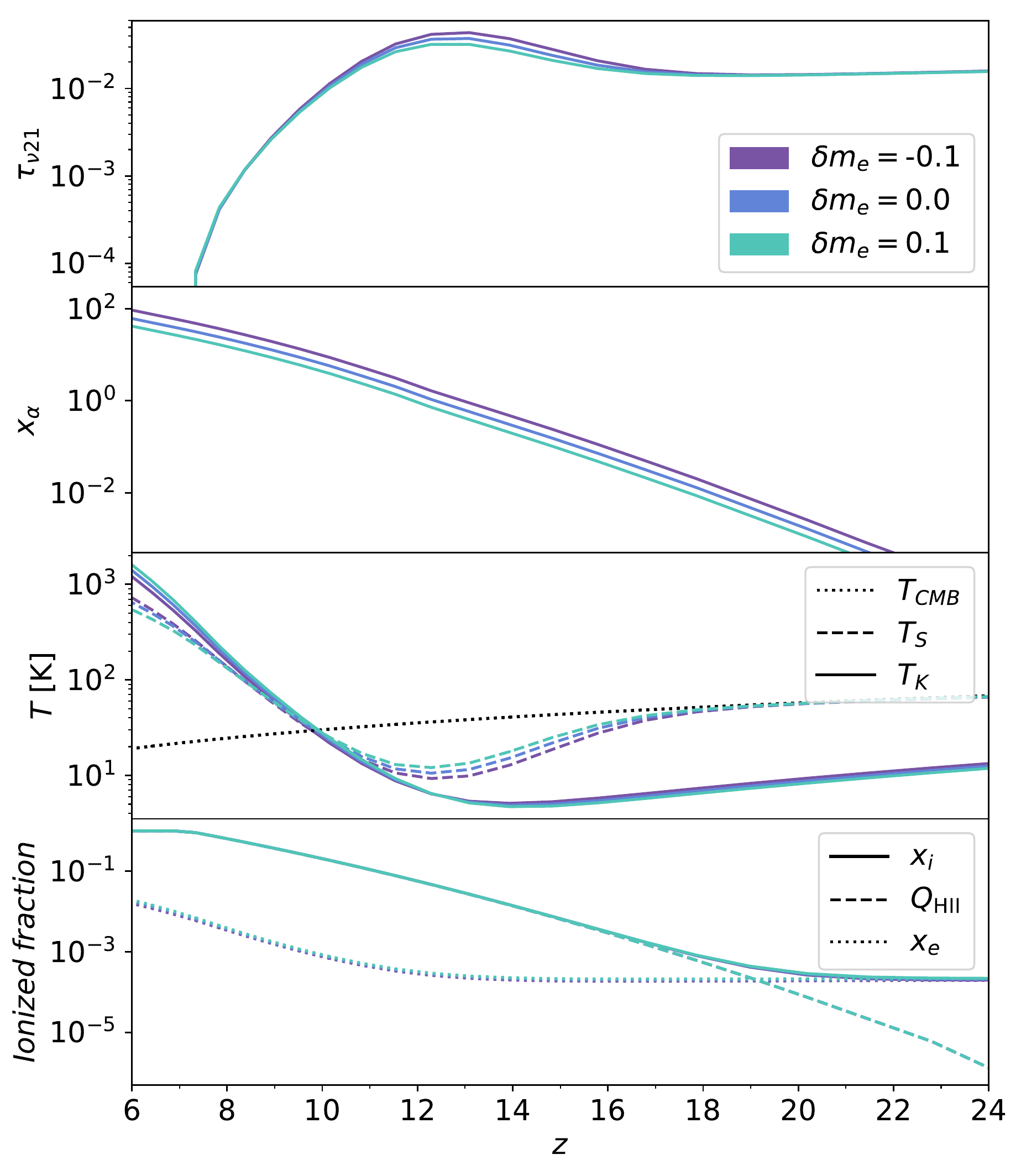}
		\caption{{Left panel: Evolution of several quantities as a function of the redshift for three values of $\delta \alpha$. In descending order: the optical depth $\tau_{\nu_{21}}$, the coupling coefficient $x_\alpha$, the kinetic, spin and CMB temperatures ($T_k$, $T_{\rm S}$ and $T_{\rm CMB}$), and the ionization fractions (the mean ionized fraction in the neutral IGM $x_e$, the volume filling factor of ionized regions $Q_{\rm H\texttt{II}}$, and the total ionization fraction $x_i$).} The astrophysical parameters have been fixed to the fiducial values in {\tt 21cmFASTv2}, $L_{{\rm X} [2 - 10~{\rm keV}]}/{\rm SFR} = 10^{40.5} \, {\rm erg} \cdot {\rm s}^{-1}M_\odot^{-1} {\rm yr}$, $N_{\rm ion} = 5 \times 10^3$, $M_{\rm turn}=5 \times 10^8 \, M_\odot$. Right panel: Same as in the left panel, but varying $\delta m_e$.}
		\label{fig:tauxalpha}
	\end{centering}
\end{figure*}

From Eqs.~(\ref{eq:Tb}) and~(\ref{eq:TgamTsappr}), it should be clear that determining the dependence of the differential brightness temperature on the fundamental constants amounts to determining the dependence of $\tau_{\nu_{21}}(\alpha, m_e)$, $x_\alpha(\alpha, m_e)$ and $T_k(\alpha, m_e)$ (notice that $T_{\rm S}(\alpha, m_e)$ is determined from the dependence of $x_\alpha(\alpha, m_e)$ and $T_k(\alpha, m_e)$). All these dependences are illustrated in \Fig{fig:tauxalpha}, as a function of redshift. They were obtained using the public code {\tt 21cmFASTv2} code~\cite{Mesinger:2007pd, Mesinger:2010ne, Park:2018ljd}, adapted as detailed below. We also indicate in Tab.~\ref{tab:constants} some of the relevant quantities, and their scaling with $\alpha$ and $m_e$. We now turn our attention to discuss the dependences in all of these functions. Let us begin by considering the optical depth of the 21cm line.

\begin{table*}
	\begin{center}
		{\def\arraystretch{1.3}
			\begin{tabular}{|c|c|c| }
				\hline
				\, Constant \, & \, Symbol \, & \, Dependence \,  \\
				\hline
				Hyperfine transition frecuency & $\nu_{21}$ & $\alpha^4 \, m_e^2$  \\
				Spontaneous emission coefficient for the 21cm transition & $A_{10}$ & $\alpha^{13} \, m_e^4$ \\
				Brightness temperature factor &  $\tau_{\nu_{21}}T_{\rm S}$ &  $\alpha^{5}$ \\
				Lyman-$\alpha$ frecuency & $\nu_\alpha$ & $\alpha^2 \, m_e$ \\
				Proper Lyman-$\alpha$ intensity & $\hat{J}_{\alpha, \star}$ & $\alpha^{-2} \, m_e^{-1}$ \\ 
				Spontaneous emission coefficient for the Lyman-$\alpha$ transition & $A_{\rm Ly\alpha}$ & $\alpha^5 \, m_e$ \\ 
				Gunn-Peterson optical depth & $\tau_{\rm GP}$ & $\alpha^{-1} \, m_e^{-2}$ \\
				Lyman-$\alpha$ coupling & $x_\alpha$ & $S_{\alpha} \, \alpha^{-10} \, m_e^{-4}$ \\ 
				Recombination case-B coefficient & $\alpha_{\rm B}$ &
				$\alpha^{3.4} \, m_e^{-3/2}$ \\
				Ground state energy of specie $i$ & $E_i = h\nu_i$ & $\alpha^2 \, m_e$ \\
			    Ionization cross section of specie $i$ & $\sigma_i$ & $\mathcal{G}_i \, \alpha^{-1} m_e^{-2} $ \\
				\hline
			\end{tabular}
		}
		\caption{\label{tab:constants}
                  Summary of some of the relevant quantities in the treatment of the 21cm signal, along with their dependence on the fundamental constants $\alpha$ and $m_e$.}
	\end{center}
\end{table*}

\subsection{Functional dependence of $\tau_{\nu_{21}}(\alpha, m_e)$}  

The optical depth of a 21cm photon in the IGM is given by
\begin{equation}
\tau_{\nu_{21}}(\alpha, m_e)=\frac{3}{32 \, \pi} \frac{A_{10}(\alpha, m_e) \, n_{\rm H} \, x_{\rm H\texttt{I}}(\alpha, m_e)}{\nu_{21}^2(\alpha, m_e) \, H \, T_{\rm S}(\alpha, m_e)}  ~,
\label{eq:tau}
\end{equation}
with $\nu_{21}(\alpha_0, m_{e0})$ the rest-frame 21cm signal frequency, $A_{10}(\alpha, m_e)$ the spontaneous decay rate of the 21cm transition (whose laboratory values are $\nu_{21,0}= 1420.4$~MHz and $A_{10,0}= 2.85 \times 10^{-15}$ s$^{-1}$, respectively), $n_{\rm H}$ the hydrogen number density, $x_{\rm H\texttt{I}}(\alpha, m_{e})$ the neutral hydrogen fraction, and $H$ the Hubble rate. All redshift dependent quantities in \Eq{eq:tau} are evaluated at $z=\nu_{21}(\alpha, m_{e})/\nu-1$. 

The rest frame frequency $\nu_{21}(\alpha, m_e)$ is simply determined by the energy splitting of the hyperfine levels in neutral hydrogen. This transition arises from the interaction between the magnetic moment of the nucleus (proton) and the magnetic field generated by the bound electron, and results in the following functional dependence~\cite{Griffiths2004Introduction}:
\begin{equation}
  \nu_{21}(\alpha, m_e)\propto \alpha^4 \, m_e^2 ~.
  \label{eq:nu21}
\end{equation}

The spontaneous emission coefficient from the triplet to the singlet state of the ground level, $A_{10}(\alpha, m_e)$, is given by
\begin{equation}
A_{10}(\alpha, m_e) = \frac{64 \,\pi^4}{3 \, g} \, \nu_{21}^3(\alpha, m_e) \, S_{21}(\alpha, m_e) ~,
\end{equation}
where $g = 3$ is the degeneracy factor of the triplet state, and $S_{21}(\alpha, m_e) = 3 \, \mu_B^2(\alpha, m_e)$ is the strength of the line~\cite{Ewenthesis}, with $\mu_B (\alpha, m_e) = \sqrt{4\pi \alpha} /(2 m_e)$ being the Bohr magneton. The $A_{10}$ dependence on the fundamental constant is thus given by
\begin{equation}
A_{10}(\alpha, m_e) \propto \alpha^{13} \, m_e^4 ~.
\label{eq:A10}
\end{equation}
Another term that depends on $\alpha$ and $m_e$ is the fraction of neutral hydrogen. Nevertheless, prior to reionization, $x_{\rm H\texttt{I}}(\alpha, m_e) = 1 - x_i(\alpha, m_e) \simeq 1$, since the total ionized fraction, $x_i$, remains small. On the other hand, reionization is driven by ionizing photons produced in collapsed structures, and is expected to have been a rapid process occurring around $z\sim 7-8$. Moreover, the exact details of this process are subject to large astrophysical uncertainties that far exceed the potential impact of variations of these fundamental constants. Thus, in what follows we will explicitly treat the neutral hydrogen fraction $x_{\rm H\texttt{I}}$ as being independent of both $\alpha$ and $m_e$.\footnote{As discussed in Section~\ref{sec:signal}, we keep the explicit dependence of $x_e(\alpha, m_e)$, the ionized fraction in the neutral IGM. This quantity is, however, at the level of $\sim 10^{-4}$ prior to reionization, while $x_{\rm H\texttt{I}}\sim 1$.}

Combining the functional dependences above, the optical depth for 21cm photons scales with $\alpha$ and $m_e$ as
\begin{equation}
 \tau_{\nu_{21}}(\alpha, m_e)  \propto \frac{\alpha^{5}}{T_{\rm S}(\alpha, m_e)} ~,
\label{eq:scal}
\end{equation}
where we have maintained the implicit dependence in the spin temperature, which we discuss below. The evolution of the optical depth with redshift is depicted in the top panels of \Fig{fig:tauxalpha} for different values of $\alpha$ and $m_e$. Notice the weak dependence on changes of $m_e$, which only comes implicitly through $T_{\rm S}$ during the epoch of Lyman-$\alpha$ coupling (we shall comment more on this below).

\subsection{ Functional dependence of $x_\alpha(\alpha, m_e)$}  

We now turn our attention to the functional dependence of the Lyman-$\alpha$ coupling, defined in \Eq{eq:TgamTsappr}, on $\alpha$ and $m_e$. As mentioned in the previous section, we neglect the dependence of the collisional coupling $x_c$, as its contribution is only important at redshifts higher than those studied here (or in high density environments, which contribute sub-dominantly to the 21cm signal due to the large volumetric suppression). 

The Lyman-$\alpha$ coupling is defined as
\begin{equation}
x_\alpha(\alpha, m_e) = \frac{64\,\pi^3}{27} \frac{T_{21}(\alpha, m_e)  \, f_{\alpha}}{A_{10}(\alpha, m_e) \, T_{\rm CMB}}  \, \left(\frac{\alpha}{m_e}\right) \, S_{\alpha}(\alpha, m_e) \,  \hat{J}_{\alpha} (\alpha, m_e) ~,
\label{eq:x_alpha}  
\end{equation}
where $T_{21} (\alpha, m_e)$ is the temperature associated to the hyperfine splitting ($T_{21} (\alpha, m_e) \propto \nu_{21} (\alpha, m_e) \propto \alpha^4 \, m_e^2$), $f_{\alpha}$ is the oscillator strength of the Lyman-$\alpha$ transition (which is independent of both $\alpha$ and $m_e$~\cite{Bethe1957}), $S_\alpha (\alpha, m_e)$ is an order unity correction factor which accounts for the detailed shape of the spectrum near the resonance~\cite{Chen:2003gc} (and approaches one for large temperatures), and $\hat{J}_\alpha (\alpha, m_e)$ is the differential Lyman-$\alpha$ flux (or proper intensity\footnote{It is related to the specific intensity, $J_\alpha$, as $J_\alpha = h\nu \hat{J}_\alpha$.}). 

The differential Lyman-$\alpha$ flux (number of photons per unit area, per unit time, per unit frequency and per steradian) contains two distinct contributions,
\begin{equation}
  \hat{J}_\alpha(\alpha, m_e) = \hat{J}_{\alpha, \rm {X}}(\alpha, m_e) +  \hat{J}_{\alpha, \star}(\alpha, m_e) ~,
\label{eq:Jalpha}
\end{equation}
where $\hat{J}_{\alpha, \rm{X}}$ is the flux resulting from X-ray excitations of neutral hydrogen, and $\hat{J}_{\alpha, \star}$ contains the contribution to the flux from UV photons between Lyman-$\alpha$ and the Lyman-$\alpha$ limit. The former flux is subdominant for reasonable astrophysical parameters, and thus can be safely neglected,\footnote{For the astrophysical parameters we consider $J_{\alpha, \rm {X}}/J_{\alpha, \star} \lesssim 10^{-3}$ for all redshifts of interest.} while the latter is given by
\begin{equation}
\hat{J}_{\alpha, \star}(\alpha, m_e) = \sum_{n=2}^{n_{\rm max}} \hat{J}_{\alpha, n}(\alpha, m_e) =
\frac{(1+z)^2}{4\pi} \sum_{n=2}^{n_{\rm max}} f_{\rm rec}(n)  \,\int_z^{z_{\textrm{max},n}} {\rm d}z' \,  \frac{\hat{\epsilon}_{\alpha}(\nu'_n,z'; \alpha, m_e)}{H(z')}  ~,  
\end{equation}
where the sum is truncated at $n_{\rm max} = 23$~\cite{Pritchard:2005an}, $f_{\rm rec} (n)$ is the probability of generating a Lyman-$\alpha$ photon from atomic level $n$ (recycled fraction of level $n$) and it is determined from selection rules and ratios of decay rates~\cite{Pritchard:2005an}, and thus, it is independent of $\alpha$ and $m_e$. The emission frequencies are given by $\nu'_n(\alpha, m_e)= \nu_n(\alpha, m_e)\frac{1+z'}{1+z}  $ with
\begin{equation}
\nu_n(\alpha, m_e) = \frac{4}{3} \, \nu_\alpha(\alpha, m_e) \left(1 - \frac{1}{n^2} \right) ~,
\label{eq:nun}
\end{equation}
where $\nu_{\alpha}$ the Lyman-$\alpha$ transition. The maximum emission redshift so that a photon enters the Lyman-$n$ resonance at redshift $z$ is $1 + z_{\textrm{max}, n} = (1 + z) [1 - (n+1)^{-2}]/[1 - n^{-2}]$. The comoving emissivity $\hat{\epsilon}_{\alpha}$ can be approximated as a factorizable expression, with the redshift dependence being dictated by the comoving star formation rate~\cite{Park:2018ljd} (in brackets) and the remaining dependences contained in the spectral distribution function $\varepsilon(\nu; \alpha, m_e)$,
\begin{equation}
\hat{\epsilon}_{\alpha}(\nu, z; \alpha, m_e) = \varepsilon(\nu;\alpha, m_e) \left[ \left(1 + \bar{\delta}_{\rm nl} \right) \int_0^\infty {\rm d}M_{\rm h} \, \frac{{\rm d}n}{{\rm d}M_{\rm h}} \, f_{\rm duty} \, \dot{M}_\star \right] ~, 
\end{equation} 
where $\bar{\delta}_{\rm nl}$ is the mean non-linear overdensity around a point in space-time, $dn/dM_{\rm h}$ is the conditional mass function following the prescription in Refs.~\cite{Barkana:2003qk, Barkana:2007xj} and $\dot{M}_\star $ is the star formation rate in a given halo and at a given redshift, as described in Ref.~\cite{Park:2018ljd}. The fraction of halos of a given mass which host galaxies, $f_{\rm duty}$ (the galaxy duty cycle), is approximated as~\cite{Park:2018ljd}
\begin{equation}
f_{\rm duty} (M_{\rm h})=e^{- M_{\rm turn}/M_{\rm h}} ~,
\label{eq:duty}
\end{equation}
where the turnover halo mass, $M_{\rm turn}$, is allowed to vary in the forecasts performed in this work (see Table~\ref{tab:astro}).

From this discussion, it is clear that the spectral distribution function $\varepsilon$ carries the $\alpha$ and $m_e$ dependence of the proper Lyman-$\alpha$ intensity $\hat{J}_{\alpha,\star}$. It is often approximated by a power law with different spectral indexes for each pair of consecutive atomic levels~\cite{Barkana:2004vb}, so that it can be written as~\cite{Lopez-Honorez:2018ipk}
\begin{equation}
\varepsilon(\nu) = N_n \, \frac{(\beta_n + 1) \, \nu_\alpha^{\beta_n}}{\nu_{n+1}^{\beta_n+1} - \nu_n^{\beta_n+1}} \, \left(\frac{\nu}{\nu_\alpha}\right)^{\beta_n} ~, 
\end{equation}
for $\nu_n \leq \nu \leq \nu_{n+1}$, with $N_n$ the number of photons (per baryon) emitted between the $n$ and $n+1$ atomic levels. The number of photons per baryon from the Lyman-$\alpha$ resonance to the Lyman limit ($N_\alpha = \sum_n N_n$) is proportional to the number of ionizing photons per baryon, $N_{\rm ion}$, assuming they probe the same stellar population~\cite{Ciardi:2003hg, Furlanetto:2006tf, Mirocha:2015jra}. In this work we assume the ratio used by default in the {\tt 21cmFASTv2} code~\cite{Mesinger:2007pd, Mesinger:2010ne, Park:2018ljd}, typical of Pop II stars. To roughly account for the possible different type of stars at high redshifts and for uncertainties in stellar population synthesis models, in our forecasts, we also allow to vary $N_{\rm ion}$ (see Table~\ref{tab:astro}).\footnote{Note that this would also account for varying the Lyman-$\alpha$ escape probability off the interstellar medium. As the absorption of Lyman-$\alpha$ photons depends on the amount of dust in galaxies, this is assumed to be one by default in \texttt{21cmFAST}, at redshifts beyond reionization. Actually, note that we do consider epochs for which the Lyman-$\alpha$ escape probability could have been different from one (namely the low-$z$ regime). However, this is not when Lyman-$\alpha$ pumping is relevant, but rather when X-ray heating and ionization are the dominant astrophysical processes determining the 21cm signal.}

Regardless of the different spectral indexes, given that the comoving emissivity is evaluated at $\nu_n$, defined in \Eq{eq:nun}, the scaling of the differential Lyman-$\alpha$ flux with the fundamental constants $\alpha$ and $m_e$ is 
\begin{equation}
\hat{J}_{\alpha, \star}(\alpha, m_e) \propto \varepsilon(\nu_n; \alpha, m_e) \propto \nu_\alpha^{-1} \propto \alpha^{-2}m_e^{-1} ~. 
\end{equation}

Finally, we turn our attention to the function $S_\alpha(\alpha, m_e)$, which is an order unity correction factor that accounts for modifications to the intensity of Lyman-$\alpha$ photons arising from radiative transfer effects~\cite{Chen:2003gc}. Around the Lyman-$\alpha$ resonance, transport effects substantially impact the coupling of the spin temperature to the color temperature. The {\tt 21cmFASTv2} code makes use of the fits of $S_\alpha$ and $T_{\rm c}$ from Ref.~\cite{Hirata:2005mz}. However, these expressions where computed for the measured local values of $\alpha$ and $m_e$. Since our goal is precisely to study the impact of variations of these fundamental constants, we need to revisit these calculations. In order to do so, we consider expressions that keep the dependence on these constants explicit. We assume the wing approximation, making use of the procedure and formulae derived in Ref.~\cite{Furlanetto:2006fs}. This approximation, in good agreement with the fit in Ref.~\cite{Hirata:2005mz} and with the exact computation, consists on evaluating the Voigt profile at its limit far away from the center of the resonance (thus, at the \textit{wings} of the profile), which significantly simplifies the calculation~\cite{Chuzhoy:2006au}. Under this approximation, $S_\alpha$ can be written in terms of special functions. However, for the range of temperatures of interest, a polynomial expansion can be obtained for temperatures above $\gtrsim 10$~K and for all redshifts considered $\lesssim 30$. This is accurate within $1$\% with respect to the exact calculation, and is given by~\cite{Furlanetto:2006fs}
\begin{equation}
S_\alpha \simeq 1 - \frac{4 \pi}{ 3 \sqrt{3} \, \Gamma(2/3)} \, \beta + \frac{8 \pi}{ 3 \sqrt{3} \,  \Gamma(1/3)} \, \beta^2 -\frac{4}{3} \, \beta^3 + \mathcal{O}(\beta^4) ~,
\label{eq:Sa}
\end{equation}
where
\begin{equation}
\beta = \eta' \left( \frac{3 \, a_{\rm V}}{2\pi \, \gamma'} \right)^{1/3} ~,
\end{equation}
with $a_{\rm V} = A_{\rm Ly\alpha}/(4\pi \Delta \nu_{\rm D})$ characterizing the width of the Voigt profile, $A_{\rm Ly\alpha}$ ($\propto \alpha^5 m_e$) being the Einstein coefficient from spontaneous emission for the Lyman-$\alpha$ transition, and $\Delta \nu_{\rm D}=\nu_\alpha \sqrt{2T_k/m_p}$ the Doppler broadening. The coefficients $\gamma'$ and $\eta'$ are related, respectively, to the Hubble expansion term and the recoil term in the radiative transfer equation. Taking into account spin exchange, they read~\cite{Chuzhoy:2005wv, Hirata:2005mz}
\begin{eqnarray}
\gamma' & = & \gamma\left(1 + \frac{T_{\rm se}}{T_k}\right)^{-1} ~, \\
\eta' & = & \eta\left(\frac{1 + T_{\rm se}/T_{\rm S}}{1 + T_{\rm se}/T_k}\right) - \frac{\Delta \nu_{\rm D}}{\nu_{\alpha}} ~,
\end{eqnarray}
with $\gamma^{-1} = \tau_{\rm GP}$ being the Gunn-Peterson optical depth to Lyman-$\alpha$ resonant scattering ($\tau_{\rm GP} \propto \alpha \, m_e^{-1} \, \nu_\alpha^{-1} \propto \alpha^{-1} \, m_e^{-2}$), $\eta = \nu_\alpha^2/(m_p \, \Delta \nu_{\rm D})$ being the mean normalized frequency drift per scattering from recoil~\cite{1981Ap.....17...69B}, and $T_{\rm se} = (2/9) \, T_k \, \nu_{21}^2/\Delta \nu_{\rm D}^2 \simeq 0.40~{\rm K} \, (\alpha/\alpha_0)^4 (m_e/m_{e0})^2$ being the characteristic temperature arising from the spin exchange contribution~\cite{Chuzhoy:2005wv}. Notice that $\beta$ also depends on $\alpha$ and $m_e$ in a non-trivial way via its dependence on $T_k$ (see the next subsection) and on $T_{\rm S}$.

Putting all the above together, the Lyman-$\alpha$ coupling $x_\alpha$ scales with $\alpha$ and $m_e$ as
\begin{equation}
x_\alpha(\alpha, m_e) \propto \frac{S_{\alpha}(\alpha, m_e)}{\alpha^{10} \, m_e^4} ~.
\end{equation}
Finally, at the  resonance and including spin exchange, the color temperature can be approximated by~\cite{Chuzhoy:2005wv, Furlanetto:2006fs}
\begin{equation}
T_{\rm c} \simeq T_k\left( \frac{1 + T_{\rm se}/T_k}{1 + T_{\rm se}/T_{\rm S}} \right) ~,
\end{equation}
where we have neglected the contribution from the Hubble flow term, which is small~\cite{Furlanetto:2006fs}. In principle, from this expression, one could rewrite $T_{\rm S}$ as a function of $T_k$ using \Eq{eq:TgamTsappr}, as done in Ref.~\cite{Chuzhoy:2005wv}. Nevertheless, note that $S_\alpha$ (and hence $x_\alpha$) also depends on $T_{\rm S}$, via $\eta'$, in a non-trivial way. We obtain $T_{\rm S}$ by iteration using the input from \Eq{eq:Sa} into \Eq{eq:TgamTsappr}. The effect of the variation of $\alpha$ and $m_e$ on $x_\alpha$ and $T_{\rm S}$, as a function of redshift, is depicted in \Fig{fig:tauxalpha}. Note that the $T_{\rm S}$ dependence on $\alpha$ and $m_e$, via $S_\alpha$, enters through $\nu_\alpha$, $\nu_{21}$, $\tau_{\rm GP}$, $A_{\rm Ly\alpha}$ and $T_k$. We discuss the $T_k$ dependence on $\alpha$ and $m_e$ in the next subsection.

\subsection{Functional dependence of $T_k(\alpha, m_e)$}  

The final ingredient necessary to determine the dependence of $T_{\rm S}$ on the fundamental constants is the temperature of the gas $T_k$. Unlike all of the functions discussed thus far, the gas temperature at the relevant epochs studied here must be obtained by solving the differential evolution equations for $T_k$ and the ionized fraction in the (mostly) neutral IGM, $x_e$,\footnote{{Notice that the total averaged ionization fraction $x_i = 1 - x_{\rm H\texttt{I}}$ is given by two contributions, one from the volume filling fraction of completely ionized regions ($Q_{\rm H\texttt{II}}$) and one from the ionized fraction in the mostly neutral IGM ($x_e$). The volume filling factor $Q_{\rm H\texttt{II}}$ can be computed using the excursion set formalism, and it is proportional to the number of ionizing photons per baryon $N_{\rm ion}$ times the fraction of mass colapsed in halos (see, e.g.,  Refs.~\cite{Furlanetto:2004nh, Mesinger:2010ne} for more details). In terms of these ionized fractions, $x_i = Q_{\rm H\texttt{II}} + x_e \, (1 - Q_{\rm H\texttt{II}})$.}} which read
\begin{eqnarray}
\frac{{\rm d}T_k(\alpha, m_e)}{{\rm d}t} & = & \frac{2}{3 \, (1 + x_e(\alpha, m_e))} \, Q_{\rm X}(\alpha, m_e) +\frac{2 \, T_k(\alpha, m_e)}{3 \, n_b}\frac{{\rm d}n_b}{{\rm d}t} - \frac{T_k(\alpha, m_e)}{1 + x_e(\alpha, m_e)} \frac{{\rm d}x_e(\alpha, m_e)}{{\rm d}t} ~,  
\label{eq:TK}  \\
\frac{{\rm d}x_e(\alpha, m_e)}{{\rm d}t} & = & \Lambda_{\rm ion, X}(\alpha, m_e) -   \alpha_{\rm B}(\alpha, m_e) \,  x_e(\alpha, m_e)^2 \, n_b \, f_{\rm H}  ~,
\label{eq:xe} 
\end{eqnarray}
where again we have made the dependence on $\alpha$ and $m_e$ explicit, and have omitted the dependence on the position $\mathbf{x}$ and redshift $z$. The baryon number density is $n_b$, $Q_{\rm X}$ is the X-ray heating rate,\footnote{Notice that for the redshifts of interest, neglecting Compton cooling is a good approximation.} $\alpha_{\rm B}$ is the case-B recombination coefficient,\footnote{Here we use case-B recombination to evaluate the ionized fraction in the neutral IGM without clumping factor, as in, e.g., Ref.~\cite{Mirocha:2014faa}, contrarily to the default implementation in {\tt 21cmFAST}.} $f_{\rm H}$ is the hydrogen number fraction and $\Lambda_{\rm ion, X}$ is the ionizing background from X-rays. The recombination coefficient $\alpha_{\rm B}$ contains a non-trivial dependence on $\alpha$ and $m_e$, although it has been argued that the approximate dependence can be captured by adopting the following scaling: $\alpha_{\rm B} \propto \alpha^{2(1+\xi)}m_e^{-3/2}$ with $\xi = 0.7$~\cite{Kaplinghat:1998ry} (see also Ref.~\cite{Chluba:2015gta} for a similar prescription).

Astrophysical X-ray sources determine both the heating and ionization rates, which can be approximated as
\begin{equation}
{Q_{\rm X}}(\alpha, m_e) = \int_{\nu_0}^{\infty} {\rm d}\nu \, \frac{4\pi J_{\rm X}(\alpha, m_e)}{h \nu} \, \sum_{i}(h \nu-E_{i}^{\rm th}(\alpha, m_e))f_{\rm heat}(\alpha, m_e) \, f_{i} \, x_{i} \, \sigma_{i}(\alpha, m_e) ~,
\label{eq:Xray_heating_rate}
\end{equation}

and

\begin{equation}
\Lambda_{\rm ion, X}(\alpha, m_e) = \int_{\nu_0}^{\infty} {\rm d}\nu \, \frac{4\pi J_{\rm X}(\alpha, m_e)}{h \nu} \, \sum_{i} f_{i} \, x_{i}(\alpha, m_e) \, \sigma_{i}(\alpha, m_e) \, F_{i}(\alpha, m_e) ~,
\label{eq:lambda_ion}
\end{equation}
with
\begin{equation}
F_i(\alpha, m_e) = (h\nu-E_{i}^{\rm th}(\alpha, m_e)) \, \sum_j \frac{f_{{\rm ion},j}(\alpha, m_e)}{E_{j}^{\rm th}(\alpha, m_e)}+ 1 ~.
\end{equation}
Here $i,j = \ion{H}{I},\,\ion{He}{I},\,\ion{He}{II}$ denote the atomic species, $f_i$ is the corresponding number fraction, $x_i(\alpha, m_e)$ is the ionization fraction, $\sigma_i(\alpha, m_e)$ is the ionization cross section, $E_{i}^{\rm th}(\alpha, m_e)=h\nu_i(\alpha, m_e)$ is the ionization threshold energy of species $i$. The lower bound of the frequency integral is set to $h\nu_{0} = 500$~eV~\cite{Das:2017fys}, the lowest frequency escaping into the IGM. The ionization cross section appearing in Eqs.~(\ref{eq:Xray_heating_rate}) and~(\ref{eq:lambda_ion}) has a non-trivial dependence on $\alpha$ and $m_e$. Nevertheless, one can factorize this function as $\sigma_{i} = \sigma_{0,i} \, \mathcal{G}_i(\nu/\nu_i)$, with the coefficient $\sigma_{0,i} \propto \alpha^{-1}m_e^{-2} $ and $\mathcal{G}_i$ a function that can be found in Refs.~\cite{Madau:2016jbv, Eide:2018rsz}. Note that, near threshold, $\mathcal{G}_i(\nu/\nu_i) \propto (\nu/\nu_i)^{-3}$~\cite{2010gfe..book, 1989agna.book}.  The fraction of energy going into heat and secondary ionizations of species $j$ is accounted for by $f_{\rm heat}$ and $f_{{\rm ion},j}$, which should a priori be functions of $\alpha$ and $m_e$. In {\tt 21cmFASTv2}, an estimate of $f_{\rm heat}$ and $f_{{\rm ion},j}$ is being used following the early work of Ref.~\cite{Furlanetto:2009uf}. In order to properly account for the $\alpha$ and $m_e$ dependence, one would have to  account for the propagation, redshifting, and deposition of energy in a similar manner as done in Ref.~\cite{Slatyer:2012yq}. This is a complicated process, however, which likely deserves a dedicated study in itself, and thus it is beyond the scope of this work. In what follows, we will treat these fractions as being independent of $\alpha$ and $m_e$. We note, however, that there are large uncertainties in the shape and normalization of the X-ray spectrum and thus, we expect that the effect of the variation of $\alpha$ and $m_e$ on these fractions, at least to some degree, can be absorbed in the astrophysical uncertainties. 

In Eqs.~(\ref{eq:Xray_heating_rate}) and (\ref{eq:lambda_ion}), $J_{\rm X}(\alpha, m_e)$ denotes the angle-averaged specific X-ray intensity (in units of erg s$^{-1}$ keV$^{-1}$ cm$^{-2}$ sr$^{-1}$)\footnote{Note that we use two conventions for treating the photons fluxes, the energy flux $J$, and the number flux $\hat{J}$, which can be related as $J = h\nu \hat{J}$.}, given by
\begin{eqnarray}
J_{\rm X}(\nu;\alpha, m_e) & = & \frac{(1+z)^3}{4\pi}\int_{z}^{\infty} {\rm d}z' \frac{{\rm d}t}{{\rm d}z'} \, \epsilon_{\rm X}(\nu_{\rm e}; \alpha, m_e) \, e^{-\tau_{\rm X} (\nu, z, z'; \alpha,m_e)} \nonumber \\
 & \simeq &\frac{(1+z)^3}{4\pi}\int_{z}^{\infty} {\rm d}z' \frac{{\rm d}t}{{\rm d}z'} \, \epsilon_{\rm X}(\nu_{\rm e}; \alpha, m_e) \, \theta\left[1 - \tau_{\rm X}(\nu, z, z'; \alpha, m_e)\right] ~,
\label{eq:JX}
\end{eqnarray}
where $\nu_{\rm e}=\nu(1+z')/(1+z)$ is the comoving frequency at emission, and $\tau_{\rm X}(\alpha,m_e)$ is the IGM optical depth from $z'$ to $z$ that characterizes the X-ray flux attenuation in the IGM. The $\tau_{\rm X}$ dependence on $\alpha$ and $m_e$ is determined by the ionization cross section $\sigma_{i}$  already taken into account, see e.g.~\cite{Mesinger:2010ne} for details. In the {\tt 21cmFASTv2} code~\cite{Mesinger:2007pd, Mesinger:2010ne, Park:2018ljd}, to speed-up the computation, a step function for the attenuation factor is assumed, such that all photons with optical depth $\tau_{\rm X} > 1$ are absorbed, whereas there is no absorption for photons with $\tau_{\rm X} \le 1$. This is represented by the last equality in \Eq{eq:JX}, where we have substituted the exponential factor by a Heaviside function. The comoving specific X-ray emissivity $\epsilon_{\rm X}(\alpha, m_e)$ is defined as
\begin{equation}
\label{eq:specific_emissivity}
\epsilon_{\rm X}(\nu; \alpha, m_e) = \frac{L_{\rm X}(\nu)}{{\rm SFR}} \left[ \left(1 + \bar{\delta}_{\rm nl} \right) \int_0^\infty {\rm d}M_{\rm h} \, \frac{{\rm d}n}{{\rm d}M_{\rm h}} \, f_{\rm duty} \, \dot{M}_\star \right] ~,
\end{equation}
where $L_{\rm X}/{\rm SFR}$ is the specific (differential) X-ray luminosity per unit star formation escaping the host galaxies (in units of erg s$^{-1}$ keV$^{-1}$ $M_\odot^{-1}$ yr), which follows a power law, $L_{\rm X} \propto (\nu/\nu_0)^{-\gamma_{\rm X}}$. We adopt $\gamma_{\rm X} = 1$, in agreement with the observed high-mass X-ray binary (HMXB) spectra~\cite{Sazonov:2017vtx}. Given the uncertainties in the X-ray emissivity, we allow the normalization to vary, and thus, we reabsorbe in this manner any dependence on $\alpha$ and $m_e$. This is done by considering the integrated luminosity (per SFR) over the energy band $(2 - 10~{\rm keV} )$, $L_{{\rm X} [2 - 10~{\rm keV}]}$/SFR and allowing it to vary around a central value in agreement with expectations for HMXB at $z = 10$~\cite{Madau:2016jbv} (see Table~\ref{tab:astro}). The other quantities in \Eq{eq:specific_emissivity} were described above. 

These definitions allow us to extract the dependence of the X-ray heating rate on $\alpha$ and $m_e$,
\begin{equation}
Q_{\rm X}(\alpha, m_e) \, \propto \, \sum_{i} \sigma_{0,i}(\alpha, m_e) \int_z^{\infty} {\rm d}z' \int_{{\rm max}\{\frac{\nu_{0}}{\nu_i},\frac{\nu_{\rm \tau=1}}{\nu_i}\}}^{\infty} {\rm d}y \, \frac{(y-1)}{y^2} \, \mathcal{G}_i(y; \alpha, m_e) \, f_{\rm heat} \, f_{i} \, x_{i} \, \propto \, \alpha^{-1} \,  m_e^{-2} \, \sum_{i} \mathcal{I}_i (\alpha, m_e)  ~,
\end{equation}
where we have exchanged the $\nu$ and $z'$ integrals in \Eq{eq:Xray_heating_rate} and applied the step function to the frequency integral, and we have defined $y\equiv\nu/\nu_i$ and $\mathcal{I}_i$ as
\begin{equation}
\mathcal{I}_i (\alpha, m_e) \propto \int_z^{\infty} {\rm d}z' \int_{{\rm max}\{\frac{\nu_{0}}{\nu_i},\frac{\nu_{\rm \tau=1}}{\nu_i}\}}^{\infty} {\rm d}y \, \frac{(y-1)}{y^2} \, \mathcal{G}_i(y; \alpha, m_e) \, f_{\rm heat} \, f_{i} \, x_{i}  ~,  
\end{equation}
where $\nu_{\rm \tau=1} (z, z'; \alpha, m_e)$ is the frequency for which the optical depth from $z'$ to $z$ is equal to 1. Analogously, the $\alpha$ and $m_e$ dependence of the ionization rate is
\begin{equation}
\Lambda_{\rm ion, X}(\alpha, m_e)  \propto \sum_{i} \nu_i^{-1} \sigma_{0,i}(\alpha, m_e) \,  \mathcal{I}_i'(\alpha, m_e)  \propto \alpha^{-3} \, m_e^{-3} \, \sum_i \mathcal{I}_i'(\alpha, m_e)   ~, 
\end{equation}
with 
\begin{equation}
\mathcal{I}_i'(\alpha, m_e) \propto \int_z^{\infty} {\rm d}z' \int_{{\rm max}\{\frac{\nu_{0}}{\nu_i},\frac{\nu_{\rm \tau=1}}{\nu_i}\}}^{\infty} {\rm d}y \, \frac{F_i(y; \alpha, m_e)}{y^2} \, \mathcal{G}_i(y,\alpha, m_e) \, f_{i} \, x_{i} ~. 
\end{equation}

The evolution of the kinetic temperature with redshift, together with the CMB and the spin temperature are depicted in \Fig{fig:tauxalpha}, for several values of $\alpha$ and $m_e$ and for the {\tt 21cmFASTv2} default values for the rest of the (astrophysical) parameters~\cite{Mesinger:2007pd, Mesinger:2010ne, Park:2018ljd}. Notice that the kinetic temperature cools adiabatically until the moment when the X-ray radiation is strong enough to heat the full IGM. On the other hand, the spin temperature is coupled firstly to the CMB. When the Lyman-$\alpha$ field is large enough to produce the Wouthuysen-Field effect, the spin temperature starts to couple to the kinetic temperature. The transition between $T_{\rm CMB}$ and $T_k$ is sensitive to changes of $\alpha$ and $m_e$ through the dependence of $x_\alpha$ and $T_k$. {In the bottom panels of \Fig{fig:tauxalpha}, we also show the evolution of the ionized fraction in the neutral IGM, $x_e$, for several values of $\alpha$ and $m_e$. As can be seen, $x_e$ depends stronger on $\alpha$ than on $m_e$, partly also due to the scaling of the recombination coefficient. For completeness, we also show the evolution of the volume filling factor of ionized regions, $Q_{\rm H\texttt{II}}$,  and of the  total averaged ionization fraction $x_i = Q_{\rm H\texttt{II}} + x_e \, (1 - Q_{\rm H\texttt{II}})$.  Notice that for $z \lesssim 20$, the evolution of the total ionization fraction is mostly dominated by the growth of the H\texttt{II} regions, enlarging the filling factor, which is driven by the astrophysical parameters $N_{\rm ion}$ and $M_{\rm turn}$, rather than by fundamental constants.}

\section{Features in the 21cm Signal}
\label{sec:signal}

For our numerical studies we make use of the latest version of the publicly available tool {\tt 21cmFASTv2}~\cite{Mesinger:2007pd, Mesinger:2010ne, Park:2018ljd}.\footnote{\url{https://github.com/andreimesinger/21cmFAST}} We have also made use of the recombination code {\tt Recfast++}~\cite{Seager:1999bc, Chluba:2010ca, AliHaimoud:2010ab, Hart:2017ndk},\footnote{\url{http://www.jb.man.ac.uk/~jchluba/Science/CosmoRec/Recfast++.html}} which was modified to account for the variation of the fundamental constants $\alpha$ and $m_e$ at the recombination period (see Ref.~\cite{Hart:2017ndk} for details). The use of {\tt Recfast++} is needed in order to obtain the initial conditions on $x_e$ and $T_k$ required in the {\tt 21cmFASTv2} code at $z = 30$. However, once again, we emphasize that this is a consequence of assuming a constant modification of the fundamental constants since redshift $z = 2.5\times 10^4$. As mentioned above, this effect is expected to be subdominant to those arising during the epoch of the cosmic dawn, when the direct 21cm probes studied in this work are available.

\begin{figure*}
\begin{centering}
\includegraphics[width=.49\textwidth]{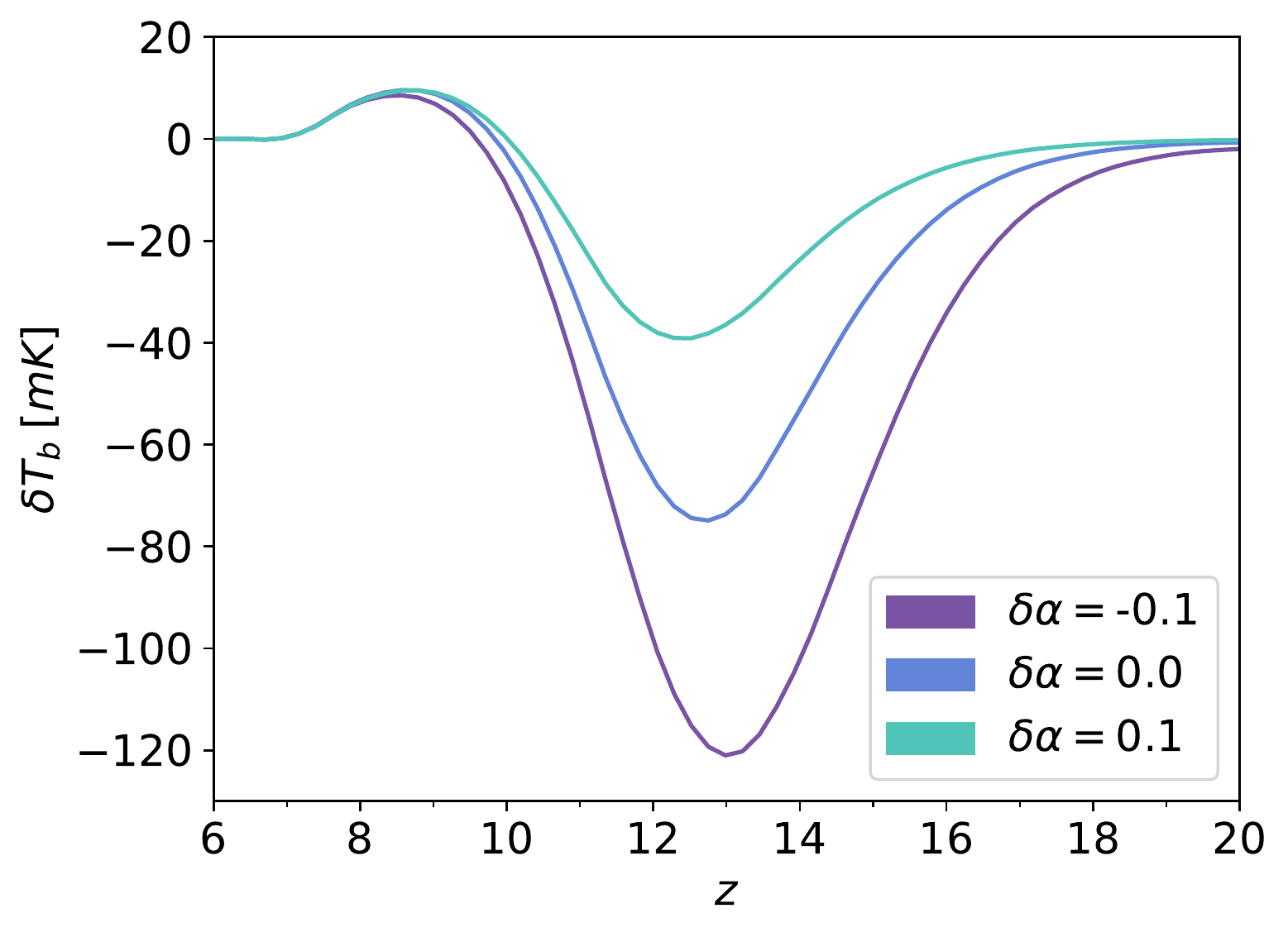}\hspace{1ex}
\includegraphics[width=.49\textwidth]{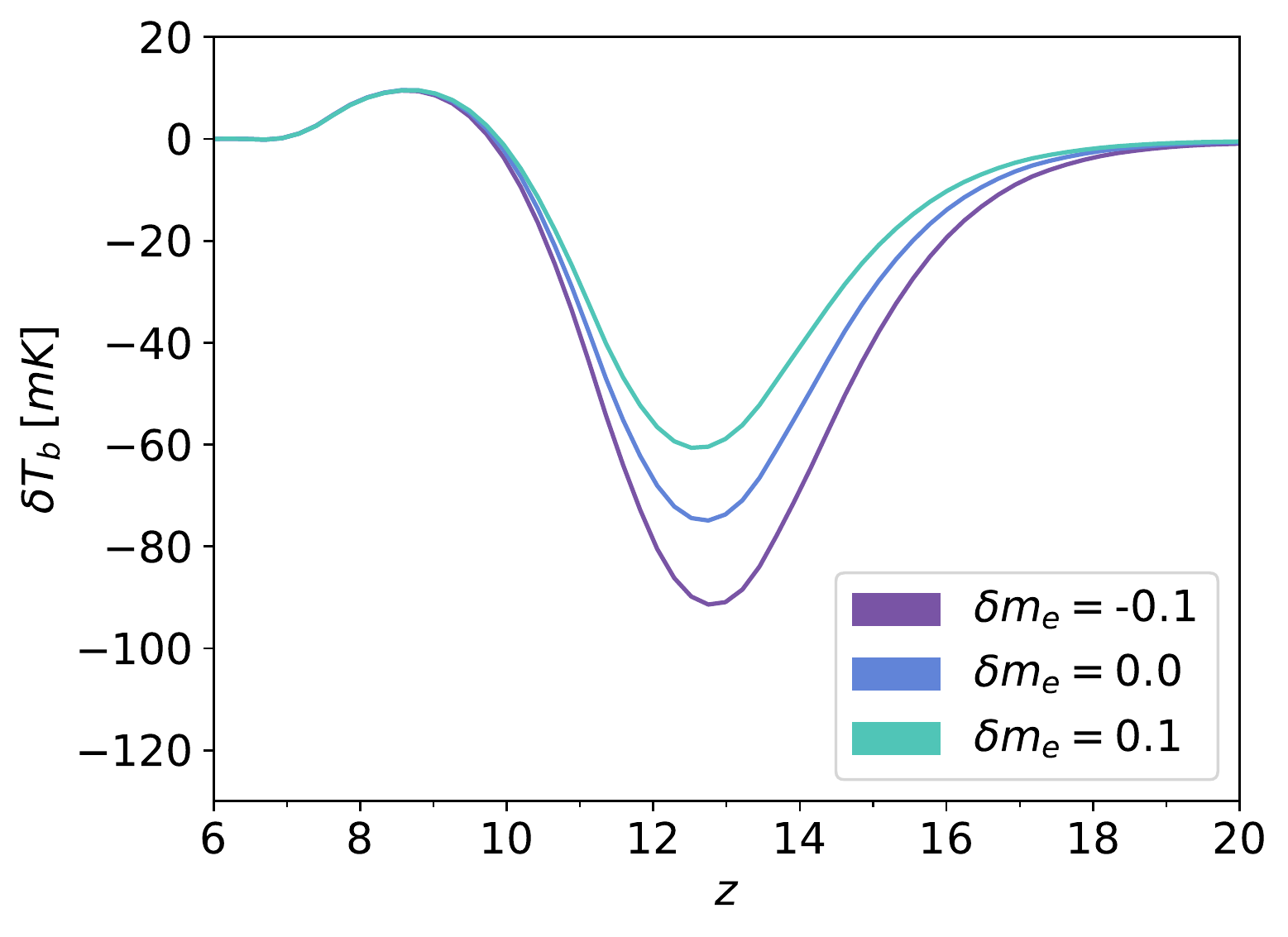}
\caption{Left panel: Differential brightness temperature evolution, as a function of the redshift, for two values of $\delta \alpha$ around the default ($\delta \alpha = 0$) case. The astrophysical parameters are the same as those in \Fig{fig:tauxalpha}. Right panel: Same as in the left panel, but varying $\delta m_e$.}
\label{fig:Tb}
\end{centering}
\end{figure*}

We show in \Fig{fig:Tb} the variation of the global differential brightness temperature $\delta T_b$ with $\alpha$ and $m_e$. The fiducial model is shown with a blue line, while the result of increasing (decreasing) $\alpha$ and $m_e$ by $10\%$ is shown as the purple/upper (green/lower) line.  Let us briefly characterize the three different regimes that can be identified in \Fig{fig:Tb} (see, e.g., Ref.~\cite{Pritchard:2011xb} for a more detailed discussion). At high redshifts near $z\sim 20$, star formation has yet to begin, and thus the Lyman-$\alpha$ flux is far too small to couple the spin temperature to the kinetic temperature of the gas. Consequently, the spin temperature is predominantly coupled to that of the CMB, and the differential brightness temperature nearly vanishes. At lower redshifts, near $z \sim 15$, the first stars have formed and produced a sufficient Lyman-$\alpha$ flux to begin coupling the spin temperature to the gas temperature, which, in this epoch, is sufficiently cooler than the CMB temperature. This produces an absorption feature in the differential brightness temperature, peaking at $z \sim 13$ for the default parameters of {\tt 21cmFASTv2}~\cite{Park:2018ljd}. This peak in the absorption spectrum is produced by the heating of the IGM, a necessary consequence of the X-ray flux associated with star formation. By $z\sim 10$, the temperature of the IGM has been heated above the temperature of the CMB photons, causing the differential brightness temperature to change from the absorption to the emission regime. Emission peaks near $z \sim 9$, and vanishes around $z \sim 7$. This arises from the fact that the differential brightness temperature is directly proportional to the neutral hydrogen fraction, which is driven to zero during the epoch of reionization.  

As can be seen in the left panel of \Fig{fig:Tb}, modifications in $\alpha$ could induce strong changes to the strength of the Lyman-$\alpha$ coupling, with negative (positive) variations producing a larger (smaller) absorption feature. While the effects of variations in the fine-structure constant during the period of X-ray heating are less significant, it is interesting to note that the heating is enhanced for negative variations of $\alpha$. The right panel of \Fig{fig:Tb}, which shows variations in $m_e$, suggests a noticeably less pronounced effect than that of $\alpha$ (for equivalent \% changes in the constant). In this case, increasing (decreasing) $m_e$ produces a smaller (larger) absorption dip, with a more uniform redshift dependence.

\begin{figure*}[t]
\begin{centering}
\includegraphics[width=.49\textwidth]{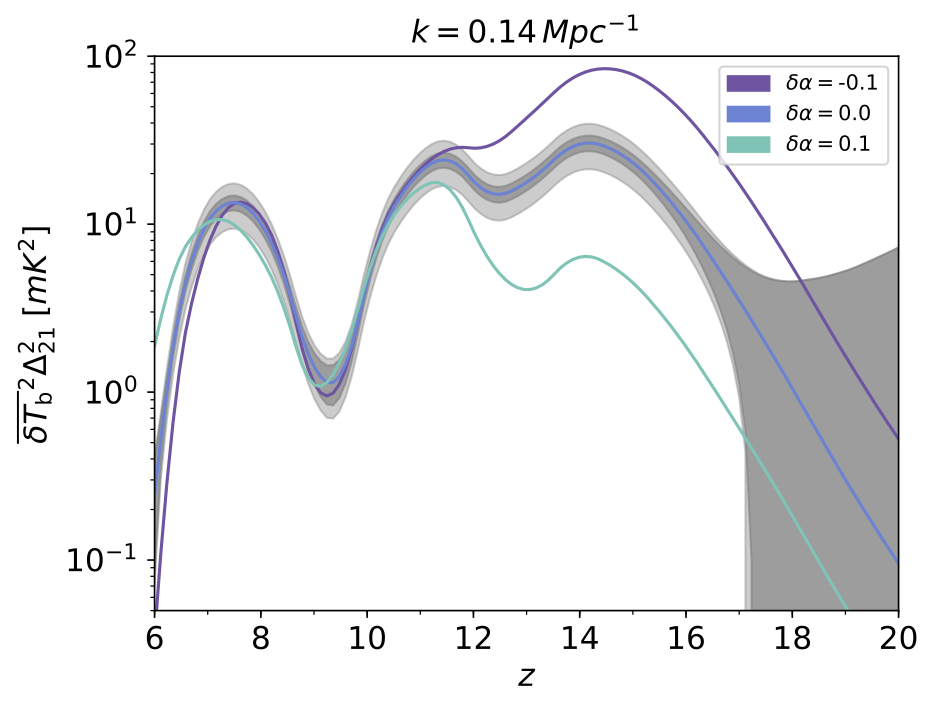}\hspace{1ex}
\includegraphics[width=.49\textwidth]{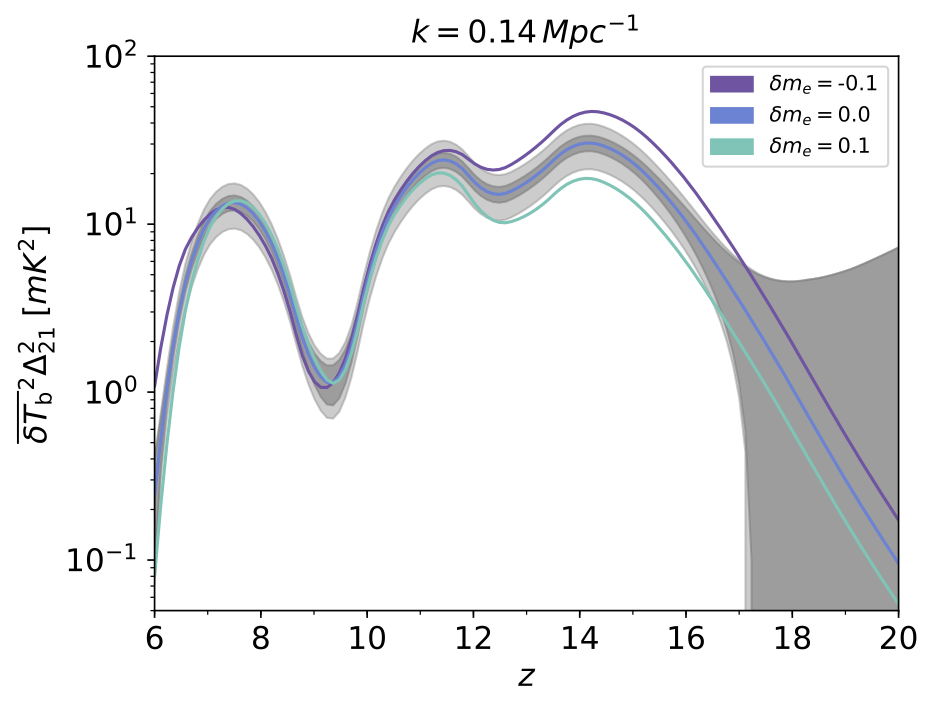} \\
\includegraphics[width=.49\textwidth]{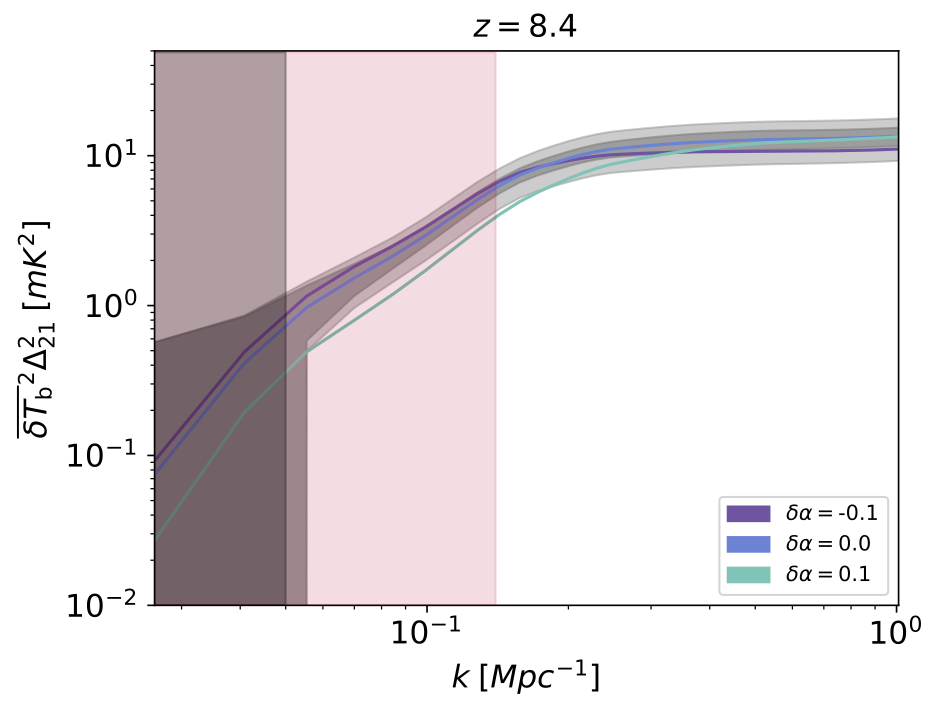}\hspace{1ex}
\includegraphics[width=.49\textwidth]{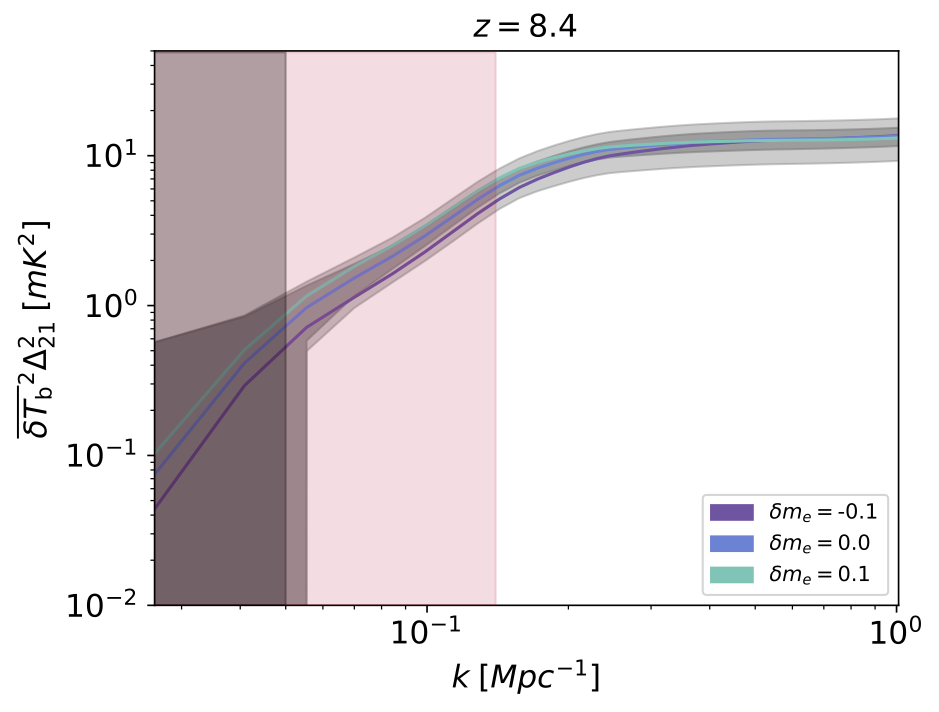}
\caption{{21cm power spectrum, as a function of redshift, for $|\delta  \alpha|\le 0.1$ (left panels) and $|\delta m_e| \le 0.1$ (right panels), with the same astrophysical parameters as in Figs.~\ref{fig:tauxalpha} and~\ref{fig:Tb}. Top panels: At a fixed scale $k = 0.14$~Mpc$^{-1}$. Bottom panels: At a fixed redshift $z = 8.4$. Dark (light) gray regions represent the total error bands with 10\% (30\%) of modeling error. Shaded areas indicate values which are not included in each of the two forecast analyses.}}	
\label{fig:P21-014}
\end{centering}
\end{figure*}

Figure~\ref{fig:P21-014} depicts the changes in the 21cm differential brightness temperature power spectrum, defined as $\overline{\delta T_b}^2(z) \Delta^2_{21} (k,z)$, for variations of $\alpha$ and $m_e$, at the scale $k = 0.14 \, {\rm Mpc}^{-1}$ {(top panels) and at fixed redshift $z = 8.4$ (bottom panels)}.  The power spectrum $\Delta^2_{21}$ is given by $\Delta^2_{21} (k,z) = (k^3/2 \pi^2) P_{21}(k,z)$, with $P_{21}(k,z)$ defined as
\begin{equation}
\langle  \widetilde{\delta}_{21} (\mathbf{k}, z)  \widetilde{ \delta}_{21}^* (\mathbf{k}^\prime, z) \rangle \equiv (2\pi)^3 \delta^{\rm D} (\mathbf{k} - \mathbf{k}^\prime) P_{21}(k,z) ~.
\label{eq:P21eq}
\end{equation}
Here, $\delta^{\rm D}$ is the Dirac delta function, the brackets indicate average quantities, and $\widetilde{\delta}_{21}(\mathbf{k}, z)$ is the Fourier transform of ${\delta}_{21}(\mathbf{x}, z) = {\delta T}_{b}(\mathbf{x}, z)/ \overline{\delta T_b}(z) - 1$. Again, one can distinguish the three different regimes (corresponding to Lyman-$\alpha$ coupling, X-ray heating, and reionization, from right to left) each of them characterized by a peak in the power spectrum. The light gray regions in \Fig{fig:P21-014}, depict the projected sensitivity of SKA, $\mathcal{S}$, obtained using {\tt 21cmSense}~\cite{Parsons:2011ew, Pober:2012zz, Pober:2013jna},\footnote{\url{https://github.com/jpober/21cmSense}} plus a modeling error $\varepsilon$ added in quadrature: $\sqrt{\mathcal{S}^2 + \varepsilon^2 \, \overline{\delta T_{\rm b}}^2 \Delta^2_{21} }$ (see Section~\ref{sec:forecasts} for more details). This modeling error attempts to account for the approximate level of disagreement between {\tt 21cmFASTv2} calculations and full hydrodynamic simulations. The default value adopted in \Fig{fig:P21-014} is $30\%$ (light gray region), although this value is somewhat \textit{adhoc} and will likely improve as the numerical methods and our understanding of astrophysics improves. Thus, we also show the results for the more optimistic case of $10\%$ (dark gray region). The scale of $k = 0.14$~Mpc$^{-1}$ has been chosen to minimize the impact of foregrounds {on the signal near the epoch of reionization~\cite{Pober:2013jna}, when the total power of astrophysical foreground emissions (e.g., diffuse galactic synchrotron emission or extragalactic point sources) are expected to be up to four orders of magnitude larger than the 21cm signal~\cite{Shaver:1999gb}. Nevertheless, signal and foregrounds contribute differently in the cylindrical 2D $k$-plane ($k_\perp, \, k_\parallel$). The two general approaches for foreground mitigation are foreground removal and foreground avoidance~\cite{Chapman:2014sfa, Chapman:2019onp}, being mode-mixing effects the main source of concern. These effects represent the contamination of the signal from the chromatic instrumental response, which can introduce spectral structure into the astrophysical foregrounds. It has been shown that in 2D $k$-space, astrophysical foregrounds are expected to be localized within a wedge-shaped region, whereas a complementary region (``EoR window'') is relatively free of foreground contamination~\cite{Datta:2010pk, Vedantham:2011mh, Morales:2012kf, Parsons:2012qh, Trott:2012md, Hazelton:2013xu, Pober:2013ig, Thyagarajan:2013eka, Jensen:2015xua}. During the last years, different foreground mitigation techniques have been developed (see, e.g., Refs.~\cite{Liu:2011hh, Dillon:2012wx, Wang:2012vn, Dillon:2013rfa, Wolz:2013wna, Liu:2014bba, Liu:2014yxa, Dillon:2015pfa, Trott:2016rjy, Kerrigan:2018rfz}). In particular, for moderate foreground models, next-generation telescope arrays could yield high-significance detections of the power spectrum for $k \sim (0.1 -1) \, h \, {\rm Mpc}^{-1}$~\cite{Pober:2013jna}.} 

Varying $\alpha$ produces a shift in all three peaks of the power spectrum (more pronounced for the Lyman-$\alpha$ peak), with negative (positive) values shifting all peaks to higher (lower) redshifts. Additionally, negative (positive) variations in $\alpha$ increase (decrease) the amplitude of each peak, as expected from the discussion of the global differential brightness temperature. Varying $m_e$ on the other hand, leaves the location of the peaks effectively unchanged, and predominantly alters only the heights, in particular at large $z$. Note that in both cases, the relative amplitudes of the peaks also change.

As stated in Section~\ref{sec:21cm}, in order to model the thermal history of the Universe, we make use of the publicly avalaible code {\tt 21cmFASTv2}~\cite{Mesinger:2007pd, Mesinger:2010ne, Park:2018ljd}.  The astrophysical parameters related to the 21cm signal included in the code and varied in this work are: the integrated X-ray luminosity (per SFR) over the energy band $(2 - 10~{\rm keV} )$, $L_{{\rm X} [2 - 10~{\rm keV}]}$/SFR; the number of ionizing photons per stellar baryon, $N_{\rm ion}$; and the halo mass characterizing the exponential decrease of star forming galaxies in halos, $M_{\rm turn}$ (such that halos with mass $M < M_{\rm turn}$ are inefficient at producing stars). The ranges in which we vary these parameters, and the fiducial values adopted for the sensitivity study, are shown in Tab.~\ref{tab:astro}. We include in Appendix~\ref{sec:astro} a brief description of how varying these parameters affects the global brightness temperature and the power spectrum (see also, e.g., Refs.~\cite{Mesinger:2012ys, Christian:2013gma, Mirocha:2015jra,  Lopez-Honorez:2016sur}).

\section{Forecasts}
\label{sec:forecasts}

Due to the computational complexity associated with obtaining the high-$z$ power spectrum, efficiently scanning the parameter space can be so computationally expensive that performing a rigorous statistical analysis becomes prohibitive. We circumvent this issue by exploiting a class of feed-forward neural networks known as multilayer perceptrons (MLPs). Specifically, for a fixed $\delta \alpha$ and $\delta m_e$, we compute the full 21cm power spectrum for $\mathcal{O}(10^4)$ randomly sampled astrophysical parameters (constrained to the range defined in Tab.~\ref{tab:astro}), and construct MLPs with seven fully connected layers, each containing $100$ hidden nodes and using leaky reLU activation functions, to emulate the output of {\tt 21cmFASTv2} for arbitrary choices of parameters. The MLPs are trained on $70\%$ of the computed power spectra, while $20\%$ was reserved for testing and $10\%$ for validation. This procedure was tested previously and was shown to be successful in producing predictions sufficiently accurate to perform a Markov chain Monte Carlo (MCMC) (see Ref.~\cite{Mena:2019nhm} for more details). For the calculations performed here, the size of typical errors induced in the log-likelihood is $\mathcal{O}(10^{-3})$.

\begin{table*}
	\begin{center}
		{\def\arraystretch{1.3}
			\begin{tabular}{|c||c|c|c| }
				\hline
				\, Parameter \, & \, $M_{\rm turn}$ [$M_\odot$] \, & \, $L_{{\rm X} [2 - 10~{\rm keV}]}$/SFR [${\rm erg} \cdot {\rm s}^{-1}M_\odot^{-1} {\rm yr}$] \, &  $N_{\rm ion}$ \\
				\hline
				Range & $10^8 - 10^9$ & $10^{40} - 10^{41}$ & \, $10^3 - 10^4$ \, \\
				Fiducial & $10^{8.5}$ & $10^{40.5}$& $10^{3.5}$ \\
				\hline
			\end{tabular}
		}
		\caption{\label{tab:astro}
			Ranges and fiducial values of the astrophysical parameters, described in Section~\ref{sec:21cm}, that are allowed to vary in the forecast runs in this work. The rest of the parameters are set to their default values in the {\tt 21cmFASTv2} code~\cite{Mesinger:2007pd, Mesinger:2010ne, Park:2018ljd}.}
	\end{center}
\end{table*}

The forecasted errors associated to future SKA measurements of the 21cm power spectrum have been estimated making use of the public code {\tt  21cmSense}~\cite{Parsons:2011ew, Pober:2012zz, Pober:2013jna} assuming the SKA System Baseline Design of Ref.~\cite{Mellema:2012ht}. In this set up, the total noise gets two type of contributions, one from thermal noise ($N$) and one from sample variance error ($S$), $\overline{\Delta}^2_{21} \equiv \overline{\delta T_b}^2 \Delta^2_{21}$. Here the thermal noise is estimated assuming an exposure of 1080~hours and a bandwidth of 8~MHz, which are the default values in {\tt 21cmSense}. Notice that using {\tt 21cmSense}, we assume a complete removal of foregrounds above  a given value of the wavenumber $k$, taken as 0.14 or  0.05 ${\rm Mpc}^{-1}$ (see below). In addition, we have chosen to add in quadrature an additional modeling error, following Ref.~\cite{Greig:2015qca}. Here we use, as default value, a rather conservative modeling error of 30\%, although we also consider the implications of a 10\% error in determining the sensitivity in our statistical analysis. Our analysis involves a multivariate Gaussian likelihood with a diagonal covariance matrix (see Ref.~\cite{Witte:2018itc} for more details). Notice that the assumption of a diagonal covariance matrix is rather optimistic, however, and may cause one to overestimate sensitivity when increasing the sampling in $z$- or  $k$-space.

\begin{figure*}[t]
	\includegraphics[width=.49\textwidth]{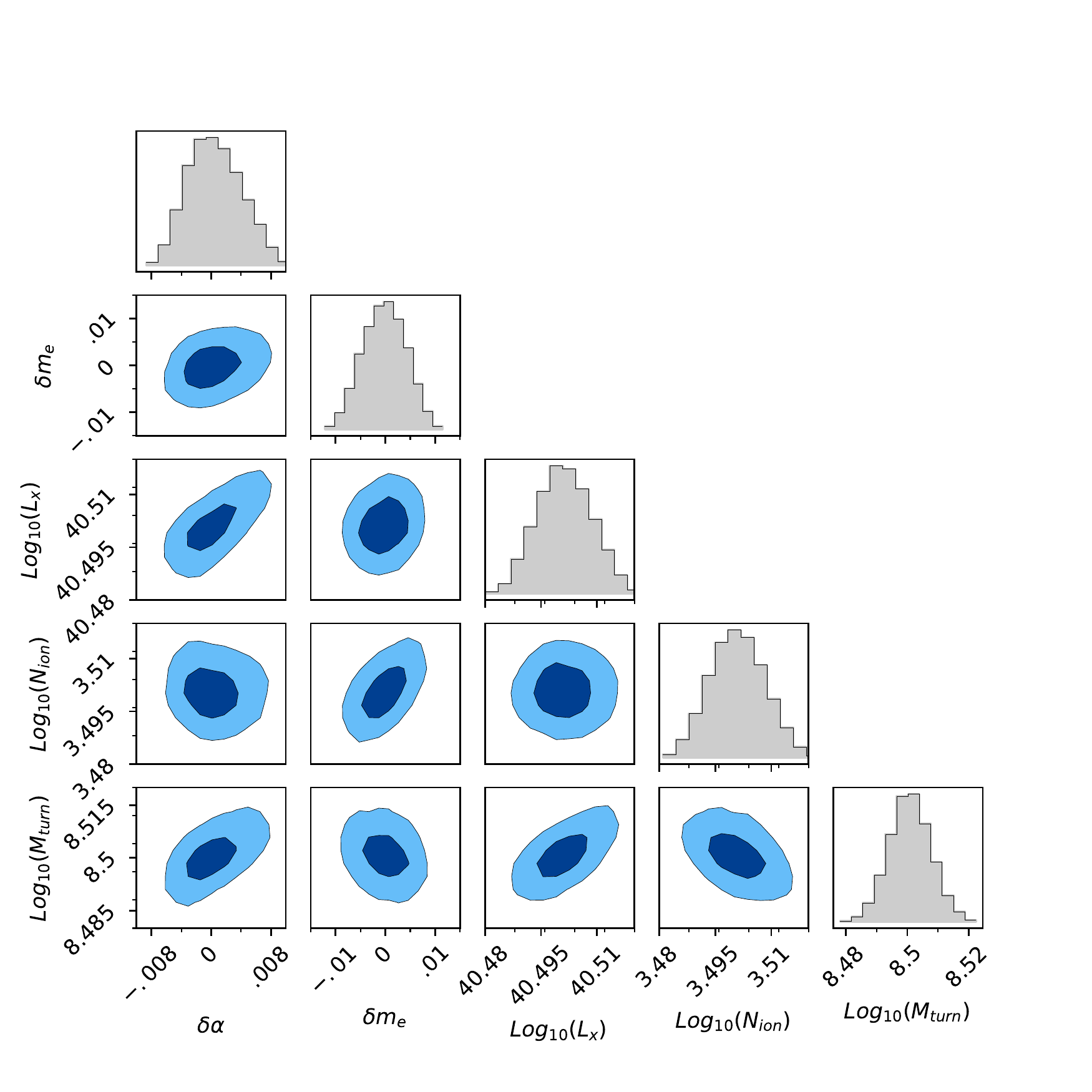}
	\includegraphics[width=.49\textwidth]{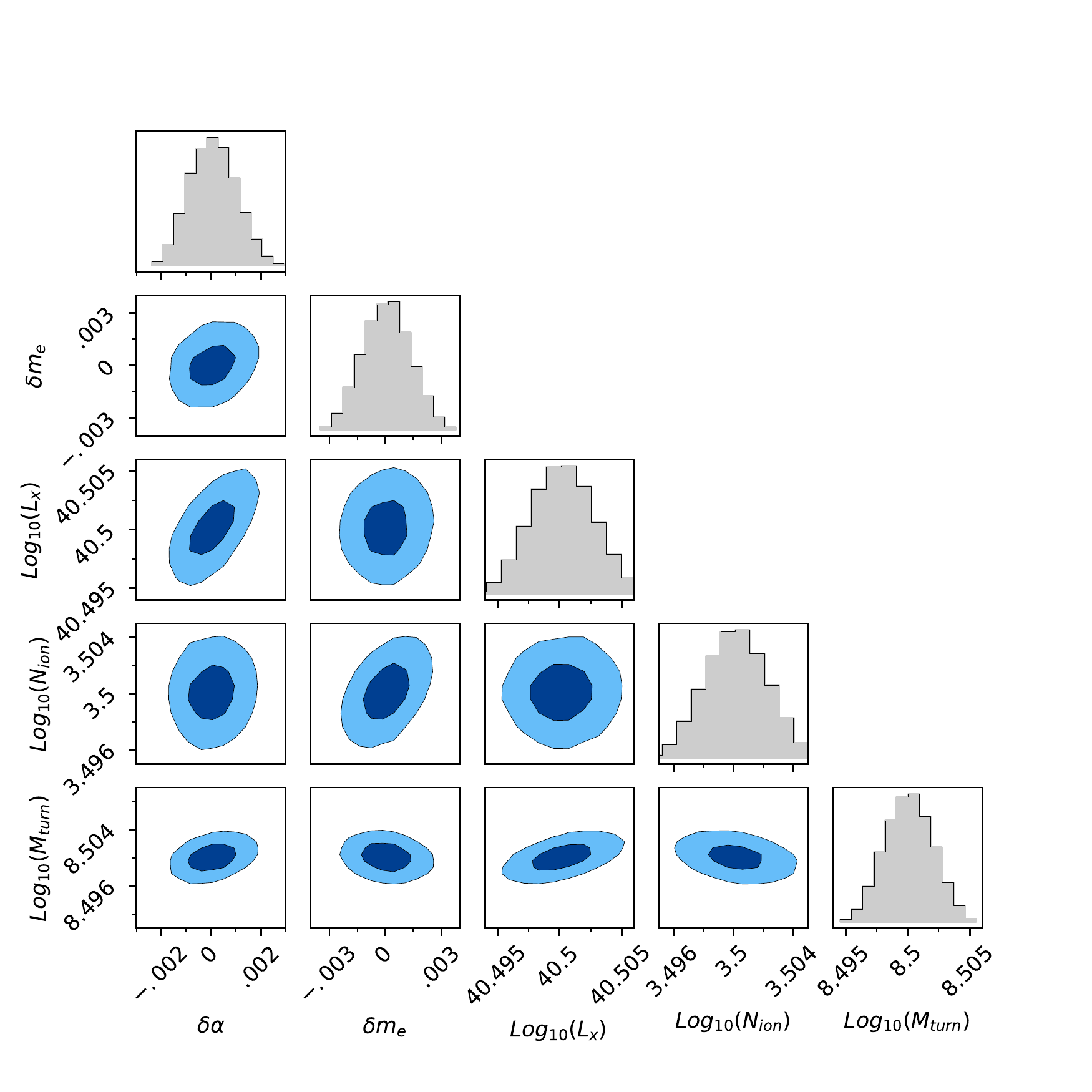}
	\caption{Expected SKA sensitivity obtained using 14 log-spaced values in redshift, ranging from $\sim 7$ to 19, and 10 (left) or 20 (right) log-spaced values in $k$, ranging from $0.14$ (left) or $0.05$ (right) to $1 \, {\rm Mpc}^{-1}$. Modeling errors are assumed to be at the $30\%$ (left) and $10\%$ (right) level. Contours are shown at 68\% and 95\% CL. Note that the scales in the left and right panels are not the same.}
	\label{fig:MCMC1}
\end{figure*}

Figure~\ref{fig:MCMC1} illustrates the expected sensitivity of SKA. We show the 68 \% and 95 \% confidence level (CL) contours for two different analyses, the difference between them being controlled by both the assumed sampling in $k$- and $z$-space, and the adopted modeling uncertainty. Note, however, that the scales used on the plots for each analysis are different. Specifically, these MCMCs assume that measurements are obtained using
\begin{enumerate}
\item 10 log-spaced values in $k$ from 0.14 to 1 ${\rm Mpc}^{-1}$, 14 log-spaced values in $z$ from $\sim 7$ to 19, and a 30\% modeling error (left panel of \Fig{fig:MCMC1});
\item 20 log-spaced values in $k$ from 0.05 to 1 ${\rm Mpc}^{-1}$, and 14 log-spaced values in $z$ from $\sim 7$ to 19, and a 10\% modeling error (right panel of \Fig{fig:MCMC1}).
\end{enumerate}

The first one can be regarded as a `conservative' analysis, while the second is more `optimistic'. The fiducial model is assumed to reside in the center of the prior space: $\log_{10}(L_{{\rm X} [2 - 10~{\rm keV}]}/({\rm SFR} \; {\rm erg} \cdot {\rm s}^{-1}M_\odot^{-1} {\rm yr}))$ = 40.5, $\log_{10}(N_{\rm ion})$~=~3.5, and $\log_{10}(M_{\rm turn}/M_\odot)$ = 8.5; and $\delta \alpha$ = 0, $\delta m_e = 0$. Some of the degeneracies can be  expected (see the discussion in Appendix~\ref{sec:astro}). In particular, focusing on the left panel of \Fig{fig:MCMC1}, we recover the expected positive degeneracies between $L_{{\rm X} [2 - 10~{\rm keV}]}$ and $M_{\rm turn}$ and between $\delta m_e$ and $N_{\rm ion}$. We note that moving from our `conservative' to `optimistic' analysis, the dominant changes that drive the increase in sensitivity are: (1) the decrease in modeling error, and (2) the shift to smaller values of $k$ (note that, if foregrounds could be avoided, the intrinsic experimental error at small $k$ is reduced). The increase in $k$-sampling alone (without the accompanying shift in $k_{\rm min}$) does not change the results dramatically, nor does the doubling of the sampling rate in redshift.

{It is interesting to notice that the constraints on $\alpha$ and $m_e$ are comparable, despite the fact that the variations in $\alpha$ produce significantly larger changes in the power spectrum, as shown in \Fig{fig:P21-014}. We have verified that the constraints obtained in \Fig{fig:MCMC1} are predominantly arising from the measurements at lower redshifts (i.e., $z \lesssim 12$), where $10\%$ variations in $\alpha$ and $m_e$ produce comparable changes in the power spectrum. While this may, at first glance, appear slightly counter-intuitive, it can be understood by noting that the small deviations across the k-space accumulate to become statistically significant, and astrophysical degeneracies in this regime are unable to mimic the deviations produced by the variations in the fundamental constants. }

The results shown in \Fig{fig:MCMC1} suggest that future 21cm experiments may be able to constrain $\alpha$ and $m_e$ at the level of $\sim {\cal O}(10^{-3})$ for redshifts $z \in [7, 19]$. For experiments like SKA, the error is expected to increase noticeably at redshifts $z \gtrsim 20$, an effect which will likely impede the ability of  ground-based interferometers to meaningfully extract limits on space-time variations of $\alpha$ or $m_e$.  {It is worth mentioning that we have performed additional analyses in which either $\alpha$ or $m_e$ were allowed to vary (rather than jointly varying both, as in \Fig{fig:MCMC1}). However, since the constraints obtained were extremely similar to those presented above, we have chosen not to include these figures. It is perhaps interesting to note that this is not the case for the CMB, where jointly varying both parameters has the effect of significantly weakening the constraint on $m_e$~\cite{Ade:2014zfo}.  Furthermore, variations of $\alpha$ have an effect in the CMB which is largely degenerate with changes in the Hubble constants $H_0$. Note that this is not the case here, as the CMB is typically far more sensitive to the cosmological parameters than 21cm cosmology~\cite{Kern:2017ccn}. }

\section{Conclusions}
\label{sec:concl}

Space-time variations in the fundamental constants of Nature are an expectation of some well-motivated theories (see, e.g., Refs.~\cite{Uzan:2002vq, Uzan:2010pm, Martins:2017yxk} for reviews). Since such variations can appear on a variety of different time scales, it is important to probe deviations in these constants from the locally measured values on both laboratory and cosmological scales, and in the early and late Universe. Current searches using the Lyman-$\alpha$ forest and from the CMB have placed stringent constraints on variations of the fine-structure constant and the electron mass at the level of $\sim 10^{-6}$ and $10^{-3}$ for redshifts $z \lesssim 6$ and $z\sim 1100$, respectively. 

Similarly to the Lyman-$\alpha$ forest, 21cm cosmology uses the redshifted absorption and emission lines of neutral hydrogen to infer astrophysical and cosmological information. High-redshift 21cm searches will soon attempt to measure the evolution of neutral hydrogen in the redshift interval $6 \lesssim z \lesssim 25$ (i.e., the epochs of reionization and cosmic dawn), exploring periods of the Universe for which we currently lack observations. Given that these radiative transitions are electromagnetic processes, the observed intensity of this signal is inherently sensitive to variations in electromagnetism, making this observational probe a particularly powerful tool to search for variations in $\alpha$ and $m_e$. 

In this work, we have detailed the complex functional dependence of the 21cm signal on the fine-structure constant $\alpha$ and on the electron mass $m_e$, and illustrated the degeneracy of varying these parameters with astrophysical uncertainties that enter the computation of the 21cm intensity. Using a 3-parameter astrophysical model that accounts for uncertainties in the intensity of X-ray and Lyman$-\alpha$ photons, as well as the onset of star formation, we show that constant variations of $\alpha$ and $m_e$ can potentially be constrained at the level of $\mathcal{O}(10^{-3})$ with future SKA measurements. This is comparable to the level obtained using observations of the CMB, but probes an epoch that is highly complementary to that of the CMB, as well as to low redshift probes from the Lyman-$\alpha$ forest.

\section*{Acknowledgments}
LLH is a Research associate of the Fonds de la Recherche Scientifique FRS-FNRS supported by the FNRS research grant number F.4520.19; by the Strategic Research Program High-Energy Physics and the Research Council of the Vrije Universiteit Brussel. OM, PVD and SJW are supported by the Spanish grant FPA2017-85985-P of the MINECO and by PROMETEO/2019/083. SPR is supported by a Ram\'on y Cajal contract, by the Spanish MINECO under grant FPA2017-84543-P, and partially, by the Portuguese FCT through the CFTP-FCT Unit 777 (UID/FIS/00777/2019). All the authors also acknowledge support from the European Union's Horizon 2020 research and innovation program under the Marie Sk\l odowska-Curie grant agreements No.\ 690575 and 674896. This work was also supported by the Spanish MINECO grant SEV-2014-0398.

\appendix

\section{Imprint of astrophysical parameters}
\label{sec:astro}

\begin{figure*}[t]
	\begin{centering}
		\includegraphics[width=.32\textwidth]{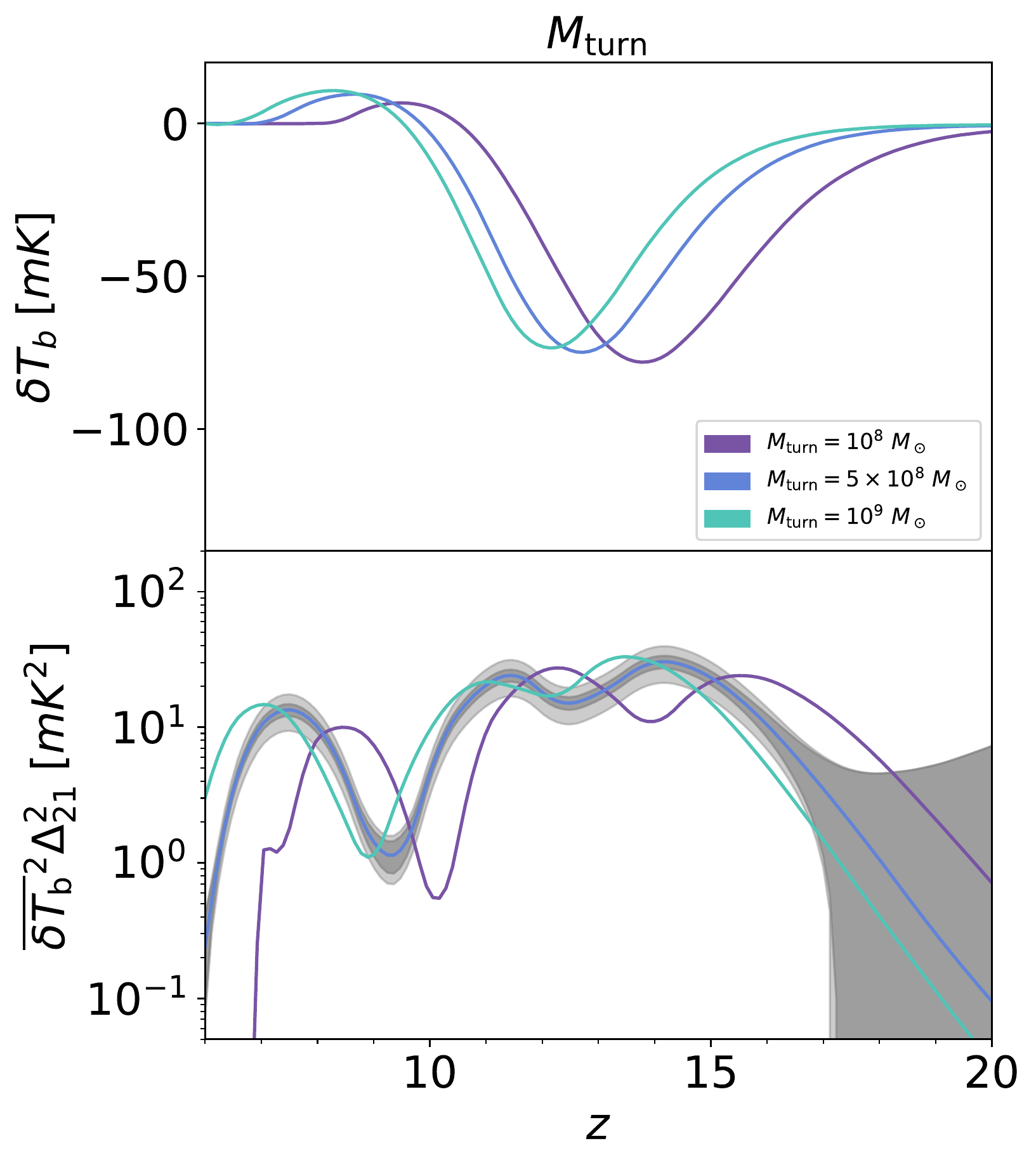}\hspace{1ex}
		\includegraphics[width=.32\textwidth]{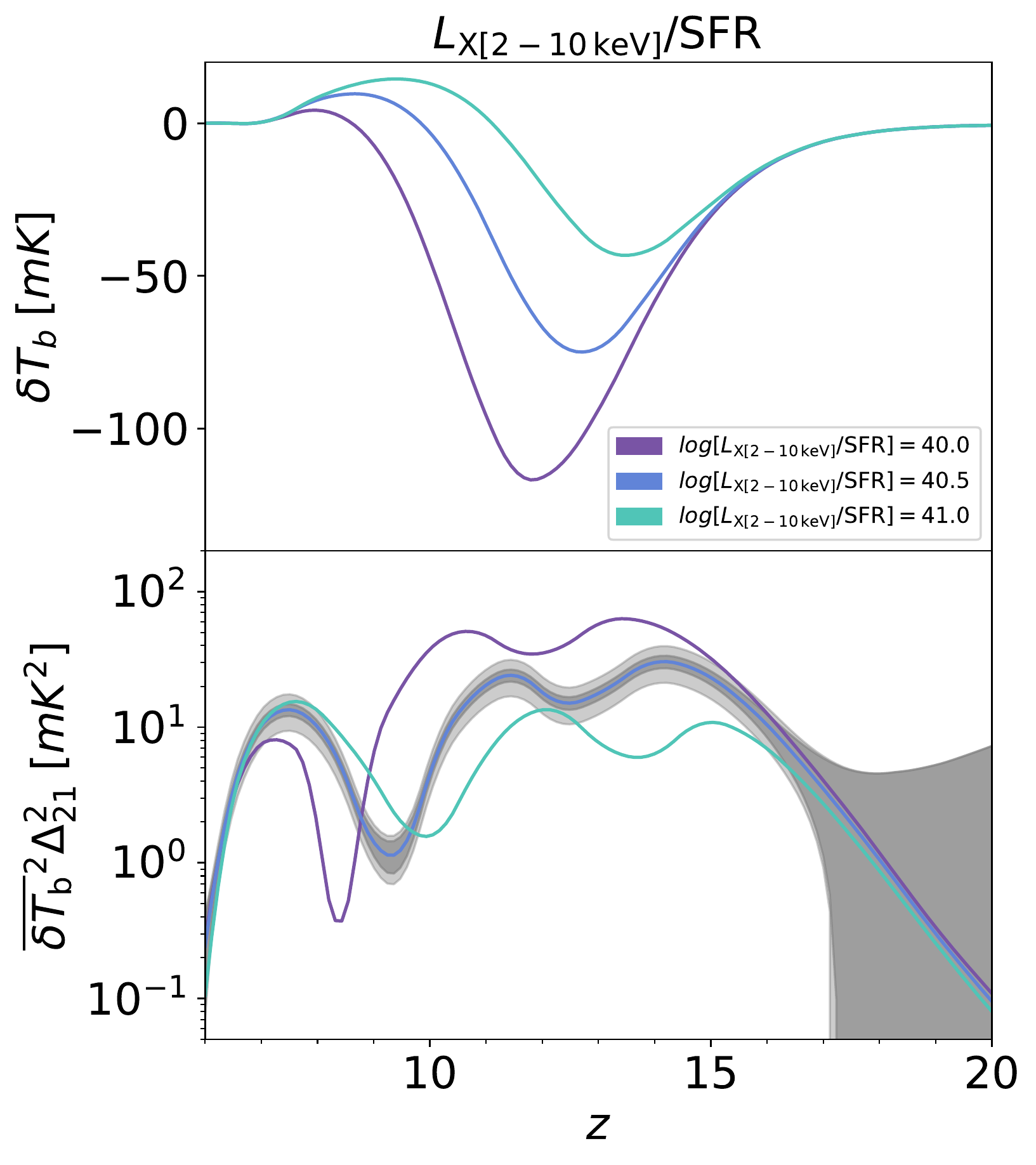}
		\includegraphics[width=.32\textwidth]{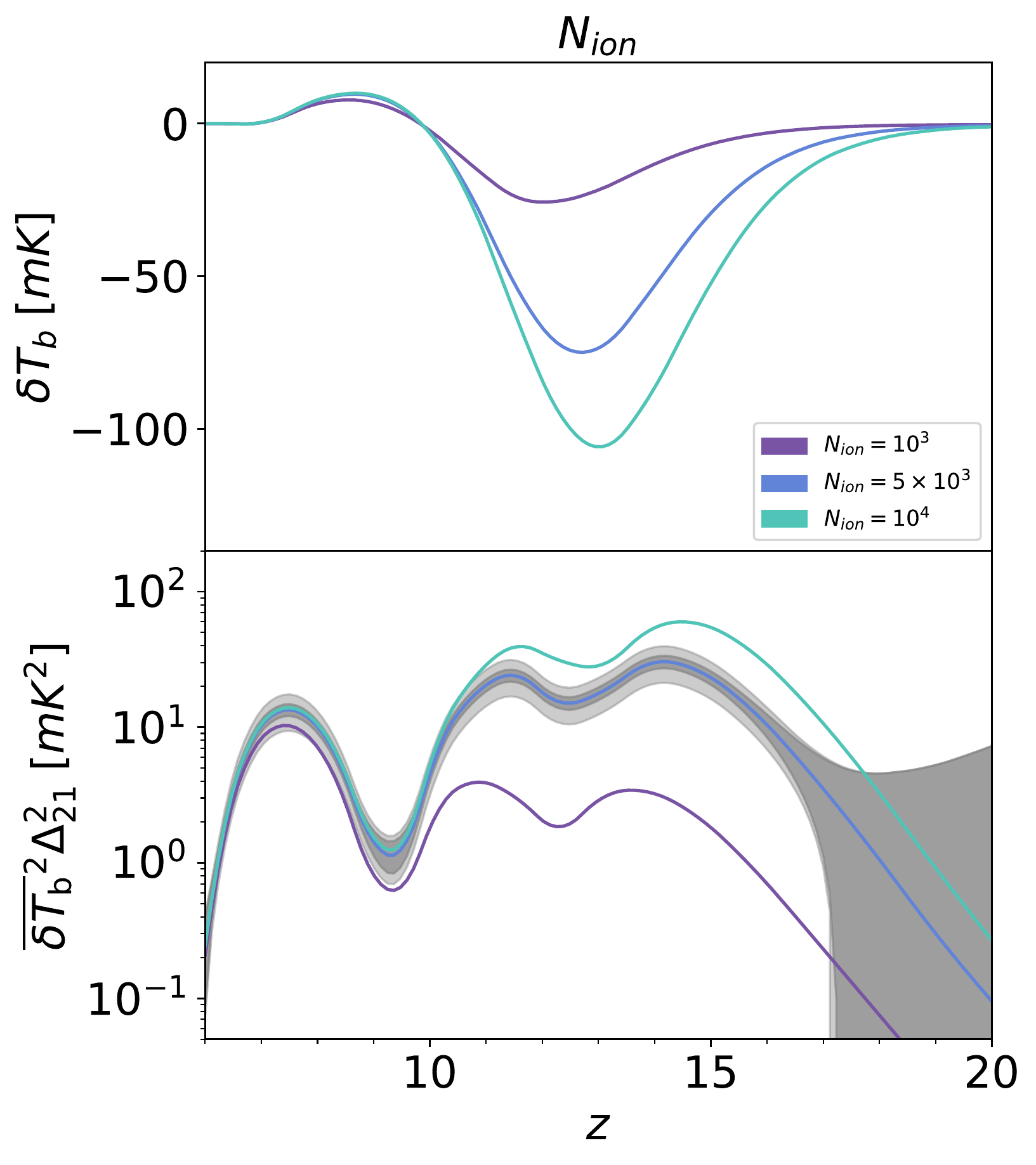}
		\caption{The 21cm differential brightness temperature (top panels) and power spectrum (bottom panels) variation with astrophysical parameters: $M_{\rm turn}$ (left panels), $L_{{\rm X} [2 - 10~{\rm keV}]}/{\rm SFR}$ (middle panels) and $N_{\rm ion}$ (right panels). Purple, blue and cyan curves stand for lower, fiducial and upper values respectively in the range considered for each parameter. The 21cm power spectrum is computed at the scale $k = 0.14$~Mpc$^{-1}$, with the band representing the SKA sensitivity.}
		\label{fig:astro}
	\end{centering}
\end{figure*}

Here, we briefly discuss the effect of varying the astrophysical parameters on the differential brightness temperature and on the 21cm power spectrum (see also, e.g., Refs.~\cite{Mesinger:2012ys, Christian:2013gma, Mirocha:2015jra,  Lopez-Honorez:2016sur}). We illustrate these effects in \Fig{fig:astro}, where blue curves correspond to the fiducial values in {\tt 21cmFASTv2}, while purple (cyan) curves show the effect of considering the minimum (maximum) value of a given astrophysical parameter.

The priors on the integrated luminosity (per SFR) over the energy band $(2 - 10~{\rm keV} )$, $L_{{\rm X} [2 - 10~{\rm keV}]}$/SFR, are indicated in Tab.~\ref{tab:astro}, and are similar to the ranges in Refs.~\cite{Mineo:2012mq, Sazonov:2017vtx, Schneider:2018xba, Monsalve:2018fno}. By varying the X-ray luminosity, we effectively modify the X-ray heating rate. Increasing $L_{{\rm X} [2 - 10~{\rm keV}]}/{\rm SFR}$ implies an earlier heating of the IGM and thus, a minimum of absorption in $\delta T_b$ and a transition to 21cm emission shifted to earlier time. Also, due to the earlier heating, a shallower absorption is expected in the background signal. The corresponding effect on the power spectrum is a shift of {\it both} the Lyman-$\alpha$ and the X-ray peak to earlier times, together with a suppression of power in both the Lyman-$\alpha$ coupling and the X-ray heating period. In contrast, the reionization peak position is left approximatively unchanged and its amplitude is increased. This is illustrated in the central plot of \Fig{fig:astro} when comparing the maximal $L_{{\rm X} [2 - 10~{\rm keV}]}$ curve (cyan) line with the fiducial one (blue line). The opposite behavior is observed considering smaller value of $L_{{\rm X} [2 - 10~{\rm keV}]}$. The effect of shifting the peak position to higher $z$ with increasing $L_{{\rm X} [2 - 10~{\rm keV}]}$ is similar to the one observed when decreasing $\alpha$. In contrast, the shallower absorption in $\delta T_b$ and the suppressed power spectrum at $z > 8$, for higher $L_{{\rm X} [2 - 10~{\rm keV}]}$, is similar to the effect of increasing $\alpha$, but also increasing $m_e$. 

The Lyman-$\alpha$ flux from stars is normalized in terms of the number of ionizing photons per stellar baryon, $N_{\rm ion}$. Assuming a Population II stars and the spectral emissivity used in \texttt{21cmFASTv2}, the normalization is of the order of $N_{\rm ion} \sim 5 \times 10^3$~\cite{Barkana:2004vb}. Increasing $N_{\rm ion}$ increases the Lyman-$\alpha$ flux at early times, leading to an earlier coupling of $T_{\rm S}$ to $T_k$. This shifts the minimum of absorption and the position of the Lyman peak, mainly to earlier times, as illustrated in the left plot of \Fig{fig:astro}. A similar effect can be obtained by decreasing $\alpha$ and $m_e$. 

The turnover halo mass, $M_{\rm turn}$, is the mass below which the abundance of active star forming galaxies is exponentially suppressed, according to the duty cycle described by \Eq{eq:duty}.\footnote{Note that $M_{\rm turn}$ plays a similar, albeit more realistic, role to the minimum virial halo mass $M_{\rm vir}^{\rm min}$ employed in other works and in previous versions of the \texttt{21cmFAST} code~\cite{Mesinger:2010ne}, which represents the limit where the exponential suppression approaches a step function.}  This affects the estimation of the Lyman-$\alpha$ and X-ray flux that shape the 21cm signal. In particular, at background level, the comoving specific X-ray emissivity is calculated from \Eq{eq:specific_emissivity}. We use a  fiducial value of $M_{\rm turn} = 5\times 10^{8} \, M_{\odot}$ and we allow the parameter to vary in the range of ${\rm \log_{10}}(M_{\rm turn}/M_\odot) = [8, \, 9]$ (see Tab.~\ref{tab:astro}). The lower limit is motivated by the atomic cooling threshold, while the upper limit is motivated by the faint end of current ultraviolet galaxy luminosity functions (see Ref.~\cite{Park:2018ljd} and references therein). Increasing $M_{\rm turn}$ suppresses the number of halos that contribute to $J_{\rm X}$ and $\hat{J}_{\alpha, *}$. As a result, the X-ray and Lyman-$\alpha$ flux are reduced at a given epoch, predominantly resulting in a redshift-dependent shift of the absorption feature. Note that a similar feature can be achieved by simultaneously varying $L_{{\rm X} [2 - 10~{\rm keV}]}$ and $N_{\rm ion}$.

\bibliography{biblio.bib}
\bibliographystyle{JHEP}

\end{document}